\documentclass[sigconf]{acmart}

\usepackage{enumitem}
\usepackage{soul, xspace}

\newcommand{\printf}{\textbf{printf}\xspace}
\AtBeginDocument{%
  \providecommand\BibTeX{{%
    \normalfont B\kern-0.5em{\scshape i\kern-0.25em b}\kern-0.8em\TeX}}}

\copyrightyear{2023}
\acmYear{2023}
\setcopyright{acmlicensed}\acmConference[CIKM '23]{Proceedings of the 32nd ACM International Conference on Information and Knowledge Management}{October 21--25, 2023}{Birmingham, United Kingdom}
\acmBooktitle{Proceedings of the 32nd ACM International Conference on Information and Knowledge Management (CIKM '23), October 21--25, 2023, Birmingham, United Kingdom}
\acmPrice{15.00}
\acmDOI{10.1145/3583780.3615012}
\acmISBN{979-8-4007-0124-5/23/10}

\begin{document}

%%
%% The "title" command has an optional parameter,
%% allowing the author to define a "short title" to be used in page headers.
\title[printf]{\textsc{printf}: Preference Modeling Based on User Reviews with Item Images and Textual Information via Graph Learning}

%%
%% The "author" command and its associated commands are used to define
%% the authors and their affiliations.
%% Of note is the shared affiliation of the first two authors, and the
%% "authornote" and "authornotemark" commands
%% used to denote shared contribution to the research.

\author{Hao-Lun Lin}
\orcid{0009-0003-3729-7266}
\affiliation{
  \institution{National Taiwan University}
  \city{Taipei}
  \country{Taiwan}
}
\email{r10944020@csie.ntu.edu.tw}

\author{Jyun-Yu Jiang}
\authornote{This work does not relate to the author's position at Amazon.}
\orcid{0000-0002-1753-8099}
\affiliation{
  \institution{Amazon Search}
  \city{Palo Alto}
  \country{CA, USA}
}
\email{jyunyu.jiang@gmail.com}

\author{Ming-Hao Juan}
\orcid{0000-0003-0203-8750}
\affiliation{
  \institution{National Taiwan University}
  \city{Taipei}
  \country{Taiwan}
}
\email{r09922083@csie.ntu.edu.tw}

\author{Pu-Jen Cheng}
\orcid{0000-0001-5892-0385}
\affiliation{
  \institution{National Taiwan University}
  \city{Taipei}
  \country{Taiwan}
}
\email{pjcheng@csie.ntu.edu.tw}

%%
%% By default, the full list of authors will be used in the page
%% headers. Often, this list is too long, and will overlap
%% other information printed in the page headers. This command allows
%% the author to define a more concise list
%% of authors' names for this purpose.
\renewcommand{\shortauthors}{Lin \emph{et al.}}

%%
%% The abstract is a short summary of the work to be presented in the
%% article.
\begin{abstract}
Nowadays, modern recommender systems usually leverage textual and visual contents as auxiliary information to predict user preference. For textual information, review texts are one of the most popular contents to model user behaviors. Nevertheless, reviews usually lose their shine when it comes to top-N recommender systems because those that solely utilize textual reviews as features struggle to adequately capture the interaction relationships between users and items. For visual one, it is usually modeled with naive convolutional networks and also hard to capture high-order relationships between users and items. Moreover, previous works did not collaboratively use both texts and images in a proper way. In this paper, we propose \printf, \underline{p}reference modeling based on user \underline{r}eviews with item \underline{i}mages a\underline{n}d \underline{t}extual in\underline{f}ormation via graph learning, to address the above challenges. Specifically, the dimension-based attention mechanism directs relations between user reviews and interacted items, allowing each dimension to contribute different importance weights to derive user representations. Extensive experiments are conducted on three publicly available datasets. The experimental results demonstrate that our proposed \printf consistently outperforms baseline methods with the relative improvements for NDCG@5 of 26.80\%, 48.65\%, and 25.74\% on Amazon-Grocery, Amazon-Tools, and Amazon-Electronics datasets, respectively. The in-depth analysis also indicates the dimensions of review representations definitely have different topics and aspects, assisting the validity of our model design.

\end{abstract}

%%
%% The code below is generated by the tool at http://dl.acm.org/ccs.cfm.
%% Please copy and paste the code instead of the example below.
%%
\begin{CCSXML}
<ccs2012>
    <concept>
        <concept_id>10002951.10003317.10003347.10003350</concept_id>
        <concept_desc>Information systems~Recommender systems</concept_desc>
        <concept_significance>500</concept_significance>
    </concept>
 </ccs2012>
\end{CCSXML}

\ccsdesc[500]{Information systems~Recommender systems}

%%
%% Keywords. The author(s) should pick words that accurately describe
%% the work being presented. Separate the keywords with commas.
\keywords{Recommender System, Collaborative Filtering, Graph Convolutional Network, Textual Content, Image Content, User Review, Dimension-based Attention}

%% A "teaser" image appears between the author and affiliation
%% information and the body of the document, and typically spans the
%% page.
% \begin{teaserfigure}
%   \includegraphics[width=\textwidth]{sampleteaser}
%   \caption{Seattle Mariners at Spring Training, 2010.}
%   \Description{Enjoying the baseball game from the third-base
%   seats. Ichiro Suzuki preparing to bat.}
%   \label{fig:teaser}
% \end{teaserfigure}

% \received{20 February 2007}
% \received[revised]{12 March 2009}
% \received[accepted]{5 June 2009}

%%
%% This command processes the author and affiliation and title
%% information and builds the first part of the formatted document.
\maketitle
\section{Introduction}

In the era of information overload, recommender systems play a crucial role in satisfying users and keeping them engaged by providing personalized recommendations. With the increasing demand for customized contents on modern E-commerce and entertainment platforms, such as Yahoo News, Amazon Shopping, and Yelp, the effectiveness of recommendations is limited by existing user-item interactions and model capacity because of data sparsity. 
To address the sparsity issues, previous studies leverage higher-order relationships to mitigate the gaps between users and items. For example, as one of the most widely adopted and successful techniques, collaborative filtering (CF)~\cite{CF} assumes low-rank relations between users and items, thereby establishing their relevance based on past interactions.
Specifically, CF assumes that users with similar preferences will also consume similar items, thereby identifying items for recommendation~\cite{AmazonCF, ItemCF, DiscreteCF, BPR, NeuralCF, HOPRec}. In recent years, graph-based models, such as knowledge graph learning ~\cite{CKE, KTUP, CKAN, KGIN, KGAT} and graph convolution networks (GCN)~\cite{PinSage, KGCN, AGCN, LightGCN}, are powerful approaches to improve the accuracy and scalability of collaborative filtering in recommendation systems by representing users and items as nodes in a graph with edges of their interactions.
In other words, the graph structure enables the capability of capturing high-order relations among users and items beyond direct interactions.
%\jjdel{ and their interactions as edges, enabling them to capture higher-order relationships between users and items beyond their direct interactions, such as common interests or preferences.} \jj{common interests kind of restricted, can discuss detailed examples later}
% \jj{ending with LightGCN is weird here. I'd suggest simply removing this sentence if you have some specific purposes. If so, you need to rephrase and elaborate it.}
% Specifically, LightGCN~\cite{LightGCN} has been shown to outperform traditional collaborative filtering methods, especially when the data is sparse or the interactions are much more complex.

% \jjrw{\jj{I don't get the point of ``Unfortunately''. The topic sentence may need to be polished.} Unfortunately, when we deep dive into the information \jj{what information? to be specific} of items and users, we found that ...
For most modern e-commerce applications, besides structured data like product categories and brands, the majority of items are also accompanied by crucial unstructured data, such as product titles and images.
Specifically, textual and image contents are the most popular unstructured data formats.
However, the potential of those unstructured data has not been fully unleashed.
For instance, conventional methods tackle texts and images with large-scale neural language models~\cite{ELMO, BERT} and convolutional neural networks~\cite{ImageNetCNN} or transformer-based visual models~\cite{ViT}, respectively, but different data formats are modeled independently so that their knowledge cannot be aligned with each other.
Moreover, to the best of our knowledge, none of the previous studies focus on simultaneously incorporating image and textual contents. 

% Although some studies like TPR~\cite{TPR} had tried to cope with the titles and descriptions as item textual contents in Top-N recommendation, it only utilizes the token-level information; nevertheless, we believe that using the global semantic of sentence-level item textual content, such as AGTM~\cite{AGTM}, could have more information gain. 
%\jjrw{\jj{too long and too casual, need to paraphrase} To be more specific, items are typically presented with eye-catching images and straightforward titles to catch users' attention, yet, most CF models, even if the graph-based ones, often neglect to use the item's feature content directly and effectively.} 
% \jjrw{\jj{``On the other hand'' is based on the implicit ``On one hand'', but I don't get the other side here. The topic sentence needs to be polished.} On the other hand, when it comes to users, privacy issues often prevent us from accessing user profiles, which poses a notable limitation for personalized recommendations.} 

Among various textual contents, user reviews can become valuable resources to model their preferences and derive user representations~\cite{ANR, AFRAM, AARM, NAML, NARRE} because detailed user opinions and item attributes can alleviate the data sparsity~\cite{shi2019deep, zheng2017joint}.
However, although some studies like RGCL~\cite{RGCL} utilize reviews as edge features in graph-based models, the relations between reviews and item contents are not considered.
This represents an unprecedented opportunity to improve the effectiveness of recommendation systems by leveraging both user reviews and item contents to generate more accurate personalized results.
% it only uses the pure CF signals for user and item representations, without considering the relationship between reviews and item contents, making the lack of explanation knowing what exact information has been propagated. 
% Despite the great progress previous works have made, there is no work trying to incorporate user reviews and item features jointly using the graph-based model in the recommendation, especially top-N tasks.

The attention mechanism~\cite{Attention} is one of the most popular ways to model user interests from interacted items~\cite{DIN}.
Previous studies like AGTM~\cite{AGTM} also utilize attention to conquer the semantic gap between users and items, further differentiating the importance of interacted items toward the user.
However, the previous attention mechanisms are based on vector similarities, where each latent dimension of item representations shares the same weight in computations.
In other words, they lose the flexibility to learn distinct similarities of diversified topics within multiple dimensions of different semantics.
Therefore, specifying different attention weights for each latent dimension can be a more suitable approach for assigning importance weights with better explanatory power.

In this paper, we present a novel framework, \underline{p}reference model based on user \underline{r}eviews with item \underline{i}mages a\underline{n}d \underline{t}extual in\underline{f}ormation via graph learning (\printf), which considers item images and textual contents as the primary source of item representations and leverages user reviews to determine user representations. Our framework consists of three sub-modules, including (1) Cross-Modality Item Modeling (CMIM), (2) Review-Aware User Modeling (RAUM), and (3) User-Item Embedding Propagation and Interaction Modeling (EPIM). CMIM fine-tunes our cross-modality feature encoder to extract image and text representations for items. RAUM leverages user reviews to distill user interests and preferences. User reviews are treated as an indicator to derive user embeddings.
Compared to previous methods, instead of using a vector-wise inner product to compute the attention scores, we propose to compute multiple dimension-wise attention scores to learn diversified topics.
EPIM employs the state-of-the-art graph convolutional network to propagate user-item embeddings and model high-order connectivity of user-item interactions.

To conclude, the main contributions of this work can be summarized as follows:
\vspace{-10pt}
\begin{itemize}
    \item To the best of our knowledge, we are the first study to innovatively incorporate unstructured contents, such as user reviews, and the images and titles of items, as supplementary sources of multi-modality information.
    \item The novel dimension-based attention technique enables us to learn varied topics within multiple embedding dimensions while modeling the connection between review embeddings and item embeddings.
    \item Our proposed framework, \printf, generates precise recommendations based on high-order relationships in user-item interactions through graph networks. Through rigorous experiments, we showcase the significance and potency of our suggested \printf, leading to the significant relative improvements for NDCG@5 of 26.80\%, 48.65\%, and 25.74\% on the Amazon-Grocery, Amazon-Tools, and Amazon-Electronics datasets, respectively.
\end{itemize}

\section{Related Work}
% In our observation, we are the first to incorporate unstructured data including user reviews and item titles and images together for Top-N recommendations; therefore, in this chapter, we would like to review some of the research topics that are related to our work.

\subsection{Graph-Based Recommendations}
In recent years, graph neural networks (GNNs) have shown promising results in various recommendation tasks. For instance, GraphRec~\cite{GraphRec} incorporates user-item interactions and item contents to learn a joint user-item representation via graph convolutional neural networks (GCNs) and propagate messages between connected nodes to predict ratings. PinSage~\cite{PinSage} uses a two-level GNN architecture to capture the structural and content-based information of items, and performs neighborhood aggregation for top-N recommendations. To further reduce the model complexity and improve the performance, LightGCN~\cite{LightGCN} is the state-of-the-art GNN-based model that minimizes the number of computational resources, only applies graph convolutional operations without non-linear activation to learn user and item embeddings. These previous studies demonstrate the effectiveness of GNNs in top-N tasks by leveraging high-order relationships successfully.

\subsection{Content-Based Recommendations}
\subsubsection{\textbf{Item Content Utilization}}
Item content features usually come in unstructured forms, such as item titles, descriptions, and images. To the best of our knowledge, the majority of studies such as TPR ~\cite{TPR} and AGTM ~\cite{AGTM} specifically focus on modeling unstructured item textual contents. TPR models token-level information between users and items by constructing an enormous knowledge graph including tokens, items, and users nodes. However, the information provided by the token-level is limited; therefore, AGTM tries to leverage transformer-based language models~\cite{BERT, Attention, SBERT} to extract item textual contents at the sentence-level. Even though both methods utilized item titles and descriptions, there is no previous model dedicated to model item textual contents and images simultaneously. However, in the field of vision-language (V-L) tasks, they have reached notable improvement and success in retrieving the information from \texttt{(image,text)} pairs.

\textbf{Vision-Language Representation Learning}
In recent years, the existing work on Vision-Language (V-L) representation learning can be classified into 3 main categories. The first category ~\cite{UNITER, CLIP} involves learning separate uni-modal encoders for images and texts and utilizing a contrastive loss to pre-train on large, noisy web data. They perform exceptionally well on image-text retrieval tasks but lack the capacity to model more intricate interactions between image and text for other V-L tasks. The second category ~\cite{VisualBERT, VILBERT} utilizes transformer-based multi-modal encoders to model interactions between image and text features, which are highly effective in downstream V-L tasks that require intricate reasoning. However, these methods often require high-resolution input images and pre-trained object detectors. Although some recent approaches aim to improve inference speed by eliminating object detectors, they result in lower performance. Therefore, hybrid models, such as ALBEF~\cite{ALBEF} and BLIP~\cite{BLIP}, have emerged as a third category, which unifies the first two categories to create robust uni-modal and multi-modal representations with superior performance on both retrieval and reasoning tasks. Furthermore, most of these recent approaches do not require object detectors, which were previously a significant bottleneck for many conventional methods. 

\vspace{-4pt}
\subsubsection{\textbf{User Review Utilization}}
Historical reviews have been widely used to improve the learning of user and item embeddings in the field of recommendation systems ~\cite{shi2019deep, zheng2017joint, ANR, AFRAM, AARM, NAML, NARRE, RGCL}. Generally speaking, the users can be represented as the reviews they have written, and the items can be represented as the reviews they have received. For instance, DeepCoNN ~\cite{DeepCoNN} utilized the TextCNN ~\cite{TextCNN} to encode features and generate user and item embeddings for rating prediction. 
% ParVecMF ~\cite{ParVecMF} employed concatenated paragraph vectors to generate user and item representations for prediction. 
In order to differentiate the importance of different reviews, the attention mechanism ~\cite{Attention} has also been introduced to improve review-based recommendation performance. 
In some studies such as NARRE~\cite{NARRE}, after the feature extraction by CNN, attention is employed to select important reviews in learning user and item representations, improving model interpretability. 
HUITA ~\cite{HUITA} designs a three-hierarchy attention mechanism to leverage word-level, sentence-level, and review-level information in user reviews. 
We can also concatenate all reviews generated by a single user into a long document, for instance, NRCA ~\cite{NRCA} integrates the document-level and review-level modeling for learning better representations. 
Last but not least, previous studies such as RGCL~\cite{RGCL} also combine the review information during the propagation of the graph-based model. It employs user reviews to interact with the corresponding users and items, and the rating-specific weights, to further get better representations for rating prediction tasks. 
In summary, from the aforementioned previous works, it is evident that user reviews play a crucial role in recommendation scenarios.
\section{\printf for Preference Modeling}

%In this section, we present \printf, \underline{p}reference modeling based on user \underline{r}eviews with item \underline{i}mages a\underline{n}d \underline{t}extual in\underline{f}ormation via graph learning, to model user preference for recommendation.
% We would like to introduce the model overview first, and detailed explanations will be shown in the rest of the sections. 

\subsection{Overview} % 3.1
\begin{figure}[!t]
    \centering
    \includegraphics[width=\linewidth]{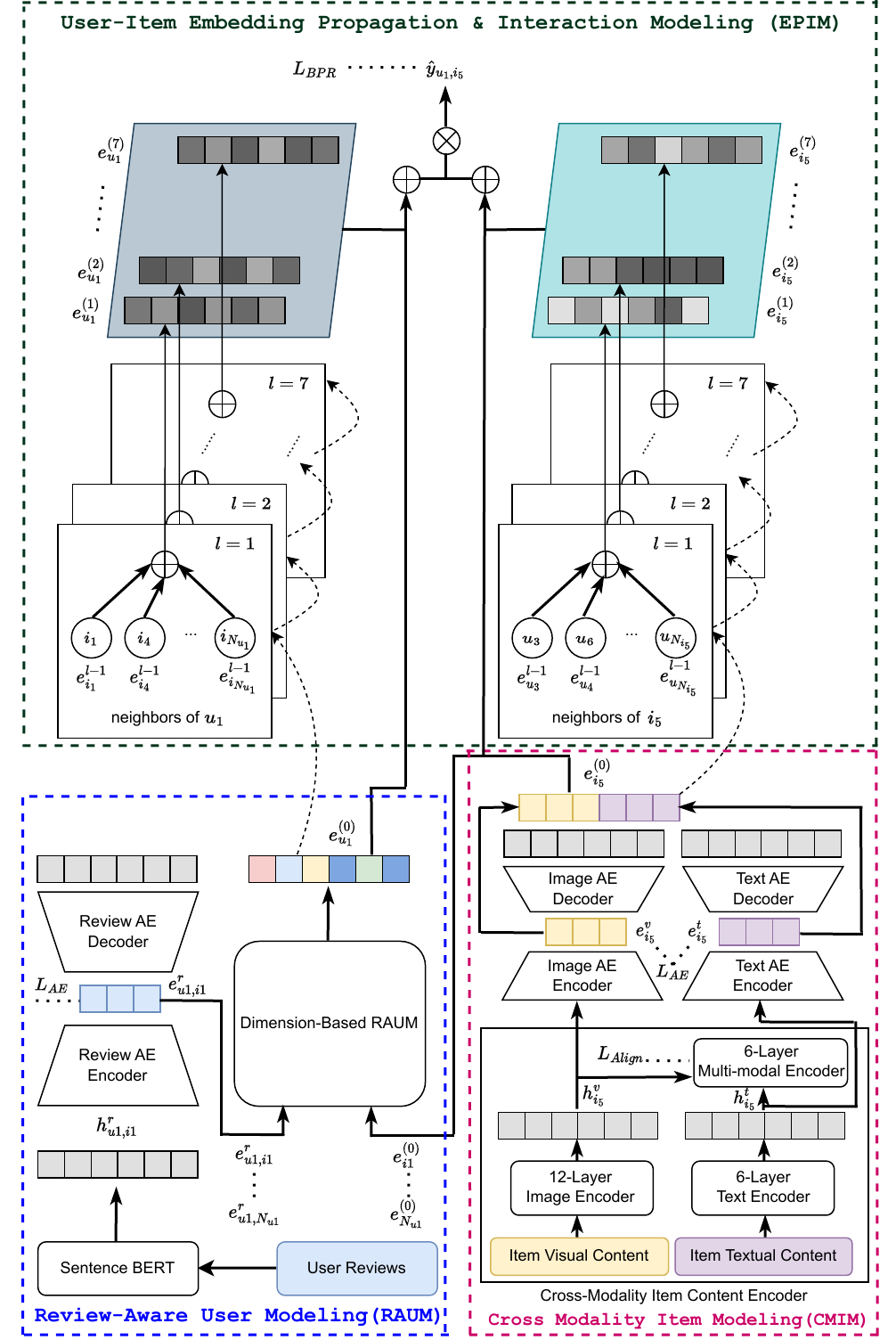}
    \caption{The overall framework of our proposed ``\printf'' for preference model.}
    \label{fig:pritnf}
    \vspace{-20pt}
\end{figure}

Referring to Figure~\ref{fig:pritnf}, \printf is composed of three main components, including (1) Cross-Modality Item Modeling (CMIM), (2) Review-Aware User Modeling (RAUM), and (3) User-Item Embedding Propagation and Interaction Modeling (EPIM). 
CMIM fine-tunes our cross-modality feature encoder to extract image and text representations for items. 
RAUM leverages user reviews to distill user interests and preferences. User reviews are treated as an indicator to derive user embeddings. 
Compared to previous methods, instead of using a vector-wise inner product operator to compute the attention scores, we propose to compute multiple dimension-wise attention scores to reflect diversified topics between user reviews and item contents. 
EPIM employs the state-of-the-art graph convolutional network to propagate user-item embeddings and model high-order connectivity of user-item interactions.

% (1) Cross-Modality Item Modeling (CMIM), which generates item embeddings from its text and image contents, (2) Review-Aware User Modeling, which generates user embeddings via review-based attention mechanisms on interacted items, conquering the issue of the semantic gap between user and item nodes, and (3) User-Item Interaction Modeling, which optimizes the generated embedding for users and items respectively. 

\noindent \textbf{Problem Statement.}
Suppose we have a user set $\mathcal{U}$, an item set $\mathcal{I}$, and the review matrix $\mathcal{R}$, where the entry $r_{u,i}$ specifies the review written by the user $u \in \mathcal{U}$ toward the item $i \in \mathcal{I}$. The observed-interactions list can be designated as $\mathcal{E} = \{(u, i, r_{u,i}) | u \in \mathcal{U}, i \in \mathcal{I}, r_{u,i} \in \mathcal{R}\}$, where each $(u, i, r_{u,i})$ tuple represents an interaction between the user $u$ and the item $i$ with the corresponding review $r_{u,i}$. Moreover, we have a set of item text contents $\mathcal{T}$, and a set of item image contents $\mathcal{M}$. For each item $i \in \mathcal{I}$, its corresponding textual and image contents are denoted as $t_i \in \mathcal{T}$ and $m_i \in \mathcal{M}$ respectively. The goal of our proposed model is to learn a function that can predict how likely a user will interact with an unseen item given interaction data $\mathcal{E}$ and item contents $\{\mathcal{T}, \mathcal{M}\}$. Note that since the review feature $r_{u,i}$ implies the positive interaction between the user $u \in \mathcal{U}$ and the item $i \in \mathcal{I}$, the review feature of the given user toward the item $r_{u,i}$ will be not be fed into the model while making the prediction, i.e., inference stage. 

\subsection{Cross-Modality Item Modeling (CMIM)} \label{sec:CMIM} \label{sec:emb_compressor}
% [現在看覺得自己以前寫太多廢話，直接註解掉～] As we mentioned before, in modern recommendation scenarios, such as in Amazon and Yelp, it's common to display recommended items with their images and titles. We believe that using these meaningful features, i.e. images and text resources, as initial representations can provide a better starting position for training. To achieve this, 

\vspace{-6pt}
\noindent \subsubsection{\textbf{Cross-Modality Feature Alignment Encoder.}}
Inspired by the ALBEF~\cite{ALBEF} model, we propose to leverage the image-text contrastive loss~\cite{Contrastive} and an additional multi-modal encoder to obtain aligned image and text representations.
Specifically, a 12-layer visual transformer~\cite{ViT} and a 6-layer BERT model~\cite{BERT} derive the image and text embeddings $\mathbf{h}^{v}_{i}$ and $\mathbf{h}^{t}_{i}$.
In the learning stage, the contrastive loss will align embeddings to each other while an additional 6-layer cross-attentive multi-modal encoder will enforce their multi-modality with self-supervised learning.
The detailed loss functions can be found in Section~\ref{sec:alignment-loss}.

\noindent \subsubsection{\textbf{Item Embedding Compressor.}}
Graph networks favor using lower-dimensional embeddings to model high-order connectivity~\cite{NGCF} while higher-dimensional embeddings usually lead to severe over-fitting~\cite{qiu2019rethinking}.
For instance, both LightGCN~\cite{LightGCN} and NGCF~\cite{NGCF} employ a fixed embedding dimension size of 64 to achieve the state-of-the-art performance at the time.
However, most of the Transformer-based text and image encoders tend to derive embeddings with a higher dimension number, e.g., 768 of BERT embeddings.
To connect the dots from Transformers to GCN, here we propose an item embedding compressor based on Auto-Encoder (AE)~\cite{AE} to reduce the number of embedding dimensions.

Formally, given the original image and text embeddings $\mathbf{h}^{v}_{i}$ and $\mathbf{h}^{t}_{i}$ of an item $i \in \mathcal{I}$, we can have AEs as:
\begin{equation}
\begin{aligned}
 \mathbf{e}^{v}_{i} = \mathcal{F}^{v}_{\text{encoder}}(\mathbf{h}^v_i), \quad
 \mathbf{\hat{h}}^{v}_{i} = \mathcal{F}^{v}_{\text{decoder}}(\mathbf{e}^v_i), \\
 \mathbf{e}^{t}_{i} = \mathcal{F}^{t}_{\text{encoder}}(\mathbf{h}^t_i), \quad
 \mathbf{\hat{h}}^{t}_{i} = \mathcal{F}^{t}_{\text{decoder}}(\mathbf{e}^t_i),
\end{aligned}
\end{equation}
where $\mathbf{e}^{v}_{i} \in \mathbb{R}^{d^\prime}$ and $\mathbf{e}^{t}_{i} \in \mathbb{R}^{d^\prime}$ are the dimension reduced image and text embeddings; $d^\prime$ is the reduced embedding dimension size, which is set to 64 in this work following previous studies~\cite{LightGCN,NGCF}; $\mathcal{F}^{v}_{\text{encoder}}(\cdot)$ and $\mathcal{F}^{v}_{\text{decoder}}(\cdot)$ as the AE of image embeddings are both two 2-layer multi-layer perceptron (MLP) modules.
Similarly, two 2-layer MLP modules $\mathcal{F}^{t}_{\text{encoder}}(\cdot)$ and $\mathcal{F}^{t}_{\text{decoder}}(\cdot)$ are applied for the AE of text embeddings.

Last but not least, the embedding of the item $i$ as the input of graph networks can be derived by concatenating the dimension-reduced item image and text embeddings together as:
\begin{equation}
\begin{aligned}
    \mathbf{e}^{(0)}_{i} = \text{concat}(\mathbf{e}^{v}_{i}, \mathbf{e}^{t}_{i})
\end{aligned}
\end{equation}

\subsection{Review-Aware User Modeling (RAUM)}\label{sec:dim-attn-raum}

To preserve the information from item contents during propagation and to avoid the semantic gap between items and users, we propose utilizing user review features as auxiliary knowledge to overcome this issue. This approach can ultimately assist in modeling users better and help bridge the semantic gap between user and item embeddings via user reviews, leading to a more robust and explainable recommendation.

\noindent \subsubsection{\textbf{User Review Feature Extraction.}}
To consider the sentence-level and paragraph-level semantics, the Sentence-Transformer~\cite{SBERT} is used as an encoder to extract contextualized global semantics from user reviews.
For the review $r_{u,i}$ written by a user $u$ toward an item $i$, we can derive the contextualized representation $\mathbf{h}^{r}_{u,i}$ as the review embedding.

\noindent \subsubsection{\textbf{Review Embedding Compressor.}}
As we mentioned in section~\ref{sec:emb_compressor}, we favor downsizing the embedding dimension for better usage in the training phase. Therefore, here we employ a similar approach to do dimension reduction via an AE. The formula can be defined in this fashion:
\begin{equation}
 \mathbf{e}^{r}_{u,i} = \mathcal{F}^{r}_{\text{encoder}}(\mathbf{h}^r_{u,i}), \quad
 \mathbf{\hat{h}}^{r}_{u,i} = \mathcal{F}^{r}_{\text{decoder}}(\mathbf{e}^r_{u,i}), 
\end{equation}
where $\mathbf{e}^{r}_{u,i} \in \mathbb{R}^{d^\prime}$ is the dimension reduced review embedding; $d^\prime$ is the reduced embedding dimension size (i.e., 64 in this work); $\mathcal{F}^{r}_{\text{encoder}}(\cdot)$ and $\mathcal{F}^{r}_{\text{decoder}}(\cdot)$ as two 2-layer MLP modules are the AE of review embeddings.

\noindent \subsubsection{\textbf{Dimension-based Review-Aware User Modeling.}}
In this paper, we assume each dimension of the review embedding represents different perspectives and aspects of the user's feedback on the interacted item.
In order to establish the connection between reviews and items, we first generate the average review embeddings of items, and calculate the dimension-based relations between user reviews and the corresponding item products as:
\begin{equation}\label{eq:dim-ri}
\begin{aligned}
    \mathbf{D} = \bar{\mathcal{R}}^{T}\mathbf{e}^{(0)}_{i}
\end{aligned}
\end{equation}
where $\bar{\mathcal{R}} \in \mathbb{R}^{|\mathcal{I}| \times d}$ is the average review embedding of the item over valid reviews written by users; $\mathbf{D} \in \mathbb{R}^{d \times d}$ is the cross-relation matrix that captures the relationship between the dimensions of the review embeddings and the item embeddings. The matrix can also be treated as a projection matrix from the latent space of user reviews toward the space of item contents. Once we have the relationship matrix between reviews and items, we can determine the dimension-based attention weight of each review towards the interacted items by multiplying the relationship matrix with the user review. Eventually, we utilize this pre-computed dimension-based attention weight to determine the significance of different interacted items by the weighted sum of the initial item embeddings for initializing user representations.
\begin{equation}\label{eq:dba-u-init}
\begin{aligned}
    \mathbf{e}^{(0)}_{u} = \sum_{i \in \mathcal{N}_u} \frac{\text{exp}(\mathbf{e}^{r}_{u,i} \times \mathbf{D})}{\sum_{j \in \mathcal{N}_u} \text{exp}(\mathbf{e}^{r}_{u,j} \times \mathbf{D})} \odot \mathbf{e}^{(0)}_{i}
\end{aligned}
\end{equation}
where $\times$ denotes matrix multiplication, $\odot$ represents element-wise multiplication (dot), and $\mathbf{e}^{(0)}_{u}$ denotes the user embedding for user $u$ as the input of the graph network. 

\subsection{User-Item Embedding Propagation and Interaction Modeling (EPIM)}

For the user-item embedding propagation phase, we leverage the structure of the graph convolutional network~\cite{LightGCN} to learn user and item embeddings via pair-wise interactions.

\noindent \subsubsection{\textbf{Message Passing and Message Aggregation.}}
We empirically follow message passing and aggregation methods similar to previous works~\cite{GCMC, LightGCN} to adopt mean aggregation in each graph convolutional layer. The formula is defined as follows:
\begin{equation}\label{eq:msg-pass}
\begin{aligned}
    \mathbf{x}^{(l+1)}_{i\rightarrow u} = \frac{\mathbf{e}_i^{(l)}}{\sqrt{|\mathcal{N}_u||\mathcal{N}_i|}} , \quad
    \mathbf{x}^{(l+1)}_{u\rightarrow i} = \frac{\mathbf{e}_u^{(l)}}{\sqrt{|\mathcal{N}_u||\mathcal{N}_i|}}
\end{aligned}
\end{equation}
where $\mathbf{x}^{(l+1)}_{i\rightarrow u}$ and $\mathbf{x}^{(l+1)}_{u\rightarrow i}$ represents the message propagated from item $i$ toward user $u$ and from user $u$ toward item $i$ through the \textit{l}-th layer to \textit{l+1}-th layer, respectively. $N_u$ and $N_i$ denote the set of items interacted by user $u$ and the set of users interacted with item $i$.
\begin{equation}\label{eq:msg-agg}
\begin{aligned}
    \mathbf{e}_u^{(l+1)} = \sum_{i \in \mathcal{N}_u} \mathbf{x}^{(l+1)}_{i\rightarrow u}, \quad
    \mathbf{e}_i^{(l+1)} = \sum_{u \in \mathcal{N}_i} \mathbf{x}^{(l+1)}_{u\rightarrow i}
\end{aligned}
\end{equation}

\noindent \subsubsection{\textbf{Layer Combination.}}
After computing $L$ layers of propagation by Equation~\ref{eq:msg-pass} and~\ref{eq:msg-agg}, we would like to aggregate different high-order relationships extracted by disparate graph convolutional layers. We simply follow a similar strategy as LightGCN~\cite{LightGCN} does, using mean pooling to combine the embeddings distilled at each layer to form the final user and item representations.
\begin{equation}
    \mathbf{e}_u = \frac{1}{L}\sum_{l=0}^L \mathbf{e}_u^{(l)}, \quad \mathbf{e}_i = \frac{1}{L}\sum_{l=0}^L \mathbf{e}_i^{(l)}
\end{equation}
where $\mathbf{e}_u$ and $\mathbf{e}_i$ are the final representations of user $u$ and item $i$.

\noindent \subsubsection{\textbf{Interaction Modeling Layer.}}
For the final layer of the proposed model, %i.e., the interaction modeling layer, 
the naive inner product operation is applied to estimate a user's preference toward a target item, which can be formulated as: 
\begin{equation}
    \hat y_{ui} = \mathbf{e}_u \cdot \mathbf{e}_i
\end{equation}
where $\hat{y}_{ui}$ denotes the matching score of the user $u$ and item $i$.

\subsection{Learning and Optimization}
\label{sec:alignment-loss}
% There are several stages in our model training, the corresponding losses will be elaborated on in the following section.

To optimize \printf, we introduce the loss functions used by each component in the framework, including alignment loss, Auto-Encoder loss, and Bayesian Personalized Ranking (BPR) loss.

\noindent \subsubsection{\textbf{Alignment Loss.}}
The alignment loss is from our cross-modality feature alignment encoder in CMIM. %section~\ref{sec:CMIM}. 
It has three sub-objectives: image-text contrastive loss and masked language modeling in uni-modality encoders, as well as image-text matching applied to the multi-modal encoder.

\begin{itemize}[leftmargin=*]

\item \textbf{Image-text Contrastive Learning Loss.}
    Image-text contrastive learning is applied to the uni-modality image encoder and text encoder respectively. It aligns image and textual features, and also trains the uni-modality encoders to better understand the semantic meaning of images and texts. Specifically, a learn-able similarity function $\textbf{s}$ is utilized.
\begin{equation}
    \textbf{s} = {W}_{img}\textbf{e}^{img}_{cls} \cdot {W}_{txt}\textbf{e}^{txt}_{cls}
\end{equation}
where ${W}_{img}$ and ${W}_{txt}$ denotes the image projection layer and text projection layer as the pretraining phase respectively. 

For each \texttt{(image,text)} pair, we compute its image-to-text and text-to-image similarity with instances via the Softmax function:
\begin{equation}\label{eq:pi2t-and-pt2i}
    \hat{y}^{i2t}(\text{I}) = \text{Softmax}(\frac{\textbf{s}(I, T)}{\pi}), \quad
    \hat{y}^{t2i}(\text{T}) = \text{Softmax}(\frac{\textbf{s}(T, I)}{\pi})
\end{equation}
where $\pi$ is a trainable parameter for controlling temperature. 

Therefore, the image-text contrastive loss from each \texttt{(image,text)} can be written as follows:
\begin{align} \label{eq:Litc}
    \mathcal{L}_{itc} &= \frac{1}{2} * \text{CrossEntropyLoss}(y^{i2t}(\text{I}), \hat{y}^{i2t}(\text{I}))  \\ 
    &+ \frac{1}{2} * \text{CrossEntropyLoss}(y^{t2i}(\text{T}), \hat{y}^{t2i}(\text{T})) \nonumber
\end{align}
where $y^{i2t}(\text{I})$ and $y^{t2i}(\text{T})$ denote the ground-truth one-hot similarity, where negative pairs have a probability of 0 and the positive pair has a probability of 1, separately.

\item {\textbf{Masked Language Modeling Loss.}}
Second, following the same pretraining objective in BERT~\cite{BERT}, masked language modeling (MLM) is applied to the multi-modal encoder. The model randomly masks out the input text tokens with a probability of 15\% and replaces them with [MASK] tokens, and predicts them using both the image and the masked text. 
\begin{equation}\label{eq:Lmlm}
    \mathcal{L}_{mlm} = \text{CrossEntropyLoss}(y^{mlm}, \hat y^{mlm}(\text{I}, \text{T}_{\text{masked}}))
\end{equation}
where $y^{mlm}$ is a one-hot encoded distribution where the ground-truth token has a value of 1. $\hat y^{mlm}(\text{I}, \text{T}_{\text{masked}})$ is the model's predicted probability distribution for a masked token with given the image representation $\text{I}$ and the masked token $\text{T}_{\text{masked}}$.

\item {\textbf{Image-Text Matching Loss.}}
Third, image-text matching is employed in the multi-modal encoder, which predicts whether a \texttt{(image, text)} pair is positive or negative, i.e., matched or not matched. The output embedding of the [CLS] token from the multi-modal encoder is treated as the joint
representation of the image-text pair, and is passed into a fully-connected network ensuing with Softmax to predict a binary classification problem.
\begin{equation}\label{eq:Litm}
    \mathcal{L}_{itm} = \text{CrossEntropyLoss}(y^{itm}, \hat y^{itm}(\text{I}, \text{T}))
\end{equation}
where the $y^{itm}$ is a 2d one-hot encoded ground truth, and $\hat y^{itm}(\text{I}, \\ \text{T})$ represents the prediction output probabilities with the given \texttt{(image,text)} pair.

\end{itemize}

To conclude, the overall alignment loss can be concisely formulated as:
\begin{equation}
    \mathcal{L}_{\text{Align}} = \mathcal{L}_{itc} + \mathcal{L}_{mlm} + \mathcal{L}_{itm}
\end{equation}

\noindent \subsubsection{\textbf{Auto-Encoder Loss.}}
In our proposed model, the AEs are used to condense user review embeddings, item image embeddings, and item text embeddings. All of these AEs are applied with mean square error (MSE) loss,
\begin{equation}
    \mathcal{L}_{\text{AE}} = \text{MSELoss}(\textbf{e}, \hat{\textbf{e}}) + \lambda *|\Theta_{\text{AE}}|^2
\end{equation}
where $\textbf{e}$ and $\hat{\textbf{e}}$ denote original input embeddings and reconstructed embeddings. Note that we also add a regularization term on trainable parameters $\Theta_{\text{AE}}$ in the AE with the $L_2$ regularization coefficient $\lambda$. We have $\lambda_{ur}$, $\lambda_{ii}$, $\lambda_{it}$ of three embedding compressors for user review, item image, and item text representations, correspondingly.

\noindent \subsubsection{\textbf{BPR Loss}}
For the user-item interaction graph in EPIM, we employ Bayesian Personalized Ranking (BPR)~\cite{BPR} loss, a pairwise loss assuming the observed interactions reflect a user's preference more than the unobserved ones. The prediction score of an observed interaction should be higher than an unobserved one on it.
\begin{equation}
    \mathcal{L}_{\text{BPR}} = - \sum_{u \in \mathcal{U}}  \sum_{i \in \mathcal{N}_u} \sum_{j \notin \mathcal{N}_u} \ln \sigma (\hat y_{ui} - \hat y_{uj}) + \lambda_{bpr} * (|\text{e}_u|^2 + |\text{e}_i|^2)
\end{equation}
where $\{(i,j) | i \in \mathcal{N}_u, j \notin \mathcal{N}_u\}$ is the pair-wise training data for user $u$, $\sigma(\cdot)$ is the Sigmoid function, adn $\lambda_{bpr}$ is the $L_2$ regularization coefficient.
\section{Experiments}
\subsection{Experimental Datasets}

We apply \printf model on three real-world Amazon datasets: Grocery, Tools, and Electronics, which are all publicly accessible\footnote{\url{https://cseweb.ucsd.edu/~jmcauley/datasets/amazon_v2/}}. All of the user-item interactions are thoroughly collected in the dataset file, it also contains review comments, which can be perfectly employed in the scenario of our work. Furthermore, we take images and titles from the metadata file as input features of items.
% In our task scenario, we split the datasets into training, validation, and test sets at a 75:5:20 ratio, respectively. Note that we sample the interactions user by user, which means for interactions belonging to a specific user, we randomly sample 75\% for train, 5\% for validation, and 20\% for testing. 
In our task scenario, we sample the interactions user by user, and split the datasets 75\% for train, 5\% for validation, and 20\% for testing.
By adopting the approach that samples the interactions user by user, we can ensure that all users receive training during the training process, eliminating the possibility of a user being included in the testing set without being part of the training set. The detailed data statistics are summarized in Table~\ref{tab:dataset-stat}.

\begin{table}[!h]
\centering
%\resizebox{.8\linewidth}{!}{
\begin{tabular}[!]{c|c|c|c}
\toprule
\textbf{Dataset} & \textbf{Grocery} & \textbf{Tools} & \textbf{Electronics} \\
\midrule
K-core & 10 & 10 & 16 \\
Users & 12,464 & 19,973 & 18,119 \\
Items & 6,644 & 11,509 & 9,952 \\
Items w/text & 6,639 & 11,496 & 9,946\\
Items w/image & 6,639 & 8,769 & 8,759  \\
Reviews (Interactions) & 205,739 & 304,439 & 440,672 \\
Density & 0.00248 & 0.00132 & 0.00244 \\
\bottomrule
\end{tabular}
%}
\caption{The statistics of the experimental datasets.}
\label{tab:dataset-stat}
\vspace{-24pt}
\end{table}

\begin{table*}[t]
\resizebox{1.0\textwidth}{!}{
\begin{tabular}[t]{l|cc|cc|cc|cc|cc|cc}
\toprule
 & \multicolumn{4}{c|}{\textbf{Amazon-Grocery}} & \multicolumn{4}{c|}{\textbf{Amazon-Tools}} & \multicolumn{4}{c}{\textbf{Amazon-Electronics}} \\
 & \textbf{R@5} & \textbf{N@5} & \textbf{R@10} & \textbf{N@10} & \textbf{R@5} & \textbf{N@5} & \textbf{R@10} & \textbf{N@10} & \textbf{R@5} & \textbf{N@5} & \textbf{R@10} & \textbf{N@10} \\ 
\midrule
BPRMF & 0.0486 & 0.0738 & 0.0907 & 0.0807 & 0.0191 & 0.0178 & 0.0274 & 0.0211 & 0.0179 & 0.0198 & 0.0292 & 0.0243 \\
LightGCN & 0.0901 & 0.0956 & 0.1141 & 0.1038 & 0.0266 & 0.0254 & 0.0383 & 0.0299 & 0.0233 & 0.0250 & 0.0374 & 0.0318 \\ 
% \midrule
RGCL & 0.0033 & 0.0031 & 0.0059 & 0.0041 & 0.0023 & 0.0019 & 0.0035 & 0.0023 & 0.0044 & 0.0047 & 0.0079 & 0.0061 \\
TPR & 0.0923 & 0.0918 & 0.1228 & 0.1031 & 0.0269 & 0.0248 & 0.0405 & 0.0302 & 0.0222 & 0.0251 & 0.0362 & 0.0301 \\
AGTM & \underline{0.1001} & \underline{0.1050} & \underline{0.1337} & \underline{0.1167} & \underline{0.0336} & \underline{0.0315} & \underline{0.0519} & \underline{0.0386} & \underline{0.0309} & \underline{0.0370} & \underline{0.0498} & \underline{0.0433} \\
\printf & \textbf{0.1233} & \textbf{0.1331} & \textbf{0.1536} & \textbf{0.1426} & \textbf{0.0466} & \textbf{0.0468} & \textbf{0.0631} & \textbf{0.0530} & \textbf{0.0390} & \textbf{0.0465} & \textbf{0.0574} & \textbf{0.0522} \\
\midrule
Improvement & 23.22\% & 26.80\% & 14.86\% & 22.19\% & 38.54\% & 48.65\% & 21.74\% & 37.23\% & 26.14\% & 25.74\% & 15.23\% & 20.52\% \\
$p$-value & 3.92E-04 & 2.94E-05 & 2.28E-03 & 8.85E-05 & 2.42E-04 & 5.41E-05 & 5.56E-04 & 3.70E-05 & 9.17E-04 & 2.27E-04 & 1.89E-03 & 9.40E-04 \\ 
\bottomrule
\end{tabular}
}
%\captionsetup{justification=centering}
\caption{Top-N recommendation performance of different methods. N@5 and R@5 denote the metrics of Recall@5 and NDCG@5. Note that the best results are highlighted in bold while the improvement indicates relative gains over the best baseline method.}
\label{tab:BigTable}

\resizebox{0.99\textwidth}{!}{
\begin{tabular}[t]{l|cc|cc|cc|cc|cc|cc}
\toprule
 & \multicolumn{4}{c|}{\textbf{Amazon-Grocery}} & \multicolumn{4}{c|}{\textbf{Amazon-Tools}} & \multicolumn{4}{c}{\textbf{Amazon-Electronics}} \\
 & \textbf{R@5} & \textbf{N@5} & \textbf{R@10} & \textbf{N@10} & \textbf{R@5} & \textbf{N@5} & \textbf{R@10} & \textbf{N@10} & \textbf{R@5} & \textbf{N@5} & \textbf{R@10} & \textbf{N@10} \\ 
\midrule
\printf & \textbf{0.1233} & \textbf{0.1331} & \textbf{0.1536} & \textbf{0.1426} & \textbf{0.0466} & \textbf{0.0468} & \textbf{0.0631} & \textbf{0.0530} & \textbf{0.0390} & \textbf{0.0465} & \textbf{0.0574} & \textbf{0.0522} \\
w/o CMIM (image) & 0.0970 & 0.1025 & 0.1298 & 0.1137 & 0.0291 & 0.0273 & 0.0460 & 0.0339 & 0.0263 & 0.0310 & 0.0434 & 0.0370 \\
w/o CMIM (title) & 0.0950 & 0.1010 & 0.1256 & 0.1114 & 0.0276 & 0.0259 & 0.0441 & 0.0324 & 0.0248 & 0.0292 & 0.0411 & 0.0348 \\
w/o RAUM & 0.1139 & 0.1250 & 0.1424 & 0.1346 & 0.0422 & 0.0427 & 0.0569 & 0.0485 & 0.0376 & 0.0452 & 0.0549 & 0.0505 \\
w/o Both & 0.0901 & 0.0956 & 0.1141 & 0.1038 & 0.0266 & 0.0254 & 0.0383 & 0.0299 & 0.0233 & 0.0250 & 0.0374 & 0.0318 \\
\bottomrule
\end{tabular}
}
\captionsetup{justification=centering}
\caption{The results of the ablation study about the contribution of different components in \printf model.}
\label{tab:exp-ablation}
% \vspace{-16pt}
\end{table*}

\subsection{Experimental Settings}
\subsubsection{\textbf{Evaluation Metric}}
Two evaluation metrics are applied to verify our model performance: Recall and Normalized Discounted Cumulative Gain (NDCG). We set all the metric rankings at the top-\{5, 10\} to evaluate and prove efficiency. For those items that have no interaction with the user are treated as negative ones, and the interacted items in the test set which is not seen in the train set are treated as positive ones. In this paper, we abbreviate \textbf{R} for Recall, and \textbf{N} for NDCG, respectively.

\subsubsection{\textbf{Baseline Methods}}
We consider five baseline models for comparisons in the experiments, including (1) \textbf{BPRMF}~\cite{BPR} as a pure CF model, (2) \textbf{LightGCN}~\cite{LightGCN} as the state-of-the-art GCN-based recommendation model, (3)  \textbf{RGCL}~\cite{RGCL} as the state-of-the-art review-based rating prediction model, (4) \textbf{TPR}~\cite{TPR} incorporating item titles and descriptions at the token-level with a knowledge graph learning, and (5) \textbf{AGTM}~\cite{AGTM} leveraging LightGCN and models both item textual contents at the sentence level and high-order connectivity in the user-item graph for the top-N recommendation.
\if 0
The baseline models we compared in the experiments are listed as:
\begin{itemize}[leftmargin=*]
    \item \textbf{BPRMF}~\cite{BPR}: BPRMF is the pure-CF model in that user and item embeddings are randomly initialized without any contents information trained with Bayesian Personalized Ranking (BPR) loss.
    \item \textbf{LightGCN}~\cite{LightGCN}: LightGCN is the state-of-the-art recommendation model in GCN-based ones, modeling high-order connectivity via user-item interactions with randomly initialized embeddings.
    \item \textbf{RGCL}~\cite{RGCL}: RGCL is the state-of-the-art review-based recommendation model for rating prediction, which uses user reviews as edge features to propagate information for users and items on the graph; moreover, they use contrastive learning to improve their model performance.
    \item \textbf{TPR}~\cite{TPR}: TPR is the recommendation model that incorporates item titles and descriptions at the token-level with a knowledge graph learning.
    \item \textbf{AGTM}~\cite{AGTM}: AGTM leverages LightGCN and models both item textual contents at the sentence level and high-order connectivity in the user-item graph for the top-N recommendation.
\end{itemize}
\fi 
For TPR\footnote{\url{https://github.com/cnclabs/codes.tpr.rec}} and RGCL\footnote{\url{https://github.com/JarenceSJ/ReviewGraph}}, we directly use their official codes which are publicly accessible on GitHub. Here, we would like to explain the reasons why we tend not to showcase the review-based baselines for model comparison. If we directly transform the rating prediction tasks, such as DeepCoNN~\cite{DeepCoNN} and NARRE~\cite{NARRE}, into top-N recommendation tasks, it empirically leads to terrible performance. Even though we perform the top-N task on the state-of-the-art review-based rating prediction model, i.e., RGCL~\cite{RGCL}, the Recall and NDCG metric are less than 0.015, which are still incredibly low. Therefore, we neglect the review-based recommendation part except for RGCL.

\subsubsection{\textbf{Implementation Details}}
The \printf is mainly implemented in PyTorch. In CMIM, the 12-layer visual transformer (ViT-B/16\footnote{\url{https://dl.fbaipublicfiles.com/deit/deit_base_patch16_224-b5f2ef4d.pth}}) is adopted for the image encoder and initialized with the pre-trained weights well-trained via ImageNet-1k from DEIT~\cite{DEIT}. For the text encoder and the multi-modal encoder, the first 6-layer and the last 6-layer of the BERT ~\cite{BERT} model (bert-base-uncased\footnote{\url{https://huggingface.co/bert-base-uncased}}) with additional cross-attention layers applied respectively. In preactical, the cross-modality feature alignment encoder is initialized with the pre-trained weights\footnote{\url{https://storage.googleapis.com/sfr-pcl-data-research/ALBEF/ALBEF_4M.pth}} from ALBEF~\cite{ALBEF}. We fine-tune the cross-modality feature alignment encoder with item \texttt{(image,text)} pairs extracted from our dataset for 3 epochs using a batch size of 32 on 1 NVIDIA RTX3090 GPU. We use the AdamW optimizer with a learning rate of $10^{-4}$ and a weight decay of $10^{-2}$. In RAUM, we use the pre-trained weights\footnote{\url{https://huggingface.co/sentence-transformers/all-mpnet-base-v2}} of Sentence-Transformer for review embedding generation.
For each representation compressor, i.e., AE, in CMIM and RAUM, the AdamW ~\cite{AdamW} optimizer with a learning rate of $10^{-3}$ and a weight decay of $10^{-2}$ is applied. For the graph in EPIM, we train it using a batch size of 4096 on 1 NVIDIA RTX3090 GPU via the AdamW optimizer with a learning rate of $10^{-3}$ and a weight decay of $10^{-2}$.

\subsection{Overall Performance}
% \begin{table*}[!]
% \resizebox{1.0\textwidth}{t!}{
% \begin{tabular}[t]{l|cc|cc|cc|cc|cc|cc}
% \toprule
%  & \multicolumn{4}{c|}{\textbf{Amazon-Grocery}} & \multicolumn{4}{c|}{\textbf{Amazon-Tools}} & \multicolumn{4}{c}{\textbf{Amazon-Electronics}} \\
%  & \textbf{R@5} & \textbf{N@5} & \textbf{R@10} & \textbf{N@10} & \textbf{R@5} & \textbf{N@5} & \textbf{R@10} & \textbf{N@10} & \textbf{R@5} & \textbf{N@5} & \textbf{R@10} & \textbf{N@10} \\ 
% \midrule
% \printf & \textbf{0.1233} & \textbf{0.1331} & \textbf{0.1536} & \textbf{0.1426} & \textbf{0.0466} & \textbf{0.0468} & \textbf{0.0631} & \textbf{0.0530} & \textbf{0.0390} & \textbf{0.0465} & \textbf{0.0574} & \textbf{0.0522} \\
% w/o CMIM (image) & 0.0970 & 0.1025 & 0.1298 & 0.1137 & 0.0291 & 0.0273 & 0.0460 & 0.0339 & 0.0263 & 0.0310 & 0.0434 & 0.0370 \\
% w/o CMIM (title) & 0.0950 & 0.1010 & 0.1256 & 0.1114 & 0.0276 & 0.0259 & 0.0441 & 0.0324 & 0.0248 & 0.0292 & 0.0411 & 0.0348 \\
% w/o RAUM & 0.1139 & 0.1250 & 0.1424 & 0.1346 & 0.0422 & 0.0427 & 0.0569 & 0.0485 & 0.0376 & 0.0452 & 0.0549 & 0.0505 \\
% w/o Both & 0.0901 & 0.0956 & 0.1141 & 0.1038 & 0.0266 & 0.0254 & 0.0383 & 0.0299 & 0.0233 & 0.0250 & 0.0374 & 0.0318 \\
% \bottomrule
% \end{tabular}
% }
% \caption{Ablation Studies.}
% \label{tab:exp-ablation}
% \end{table*}

The complete main results among three Amazon datasets are summarized in Table~\ref{tab:BigTable}. According to the table, we can obtain the following observations: 

BPRMF is a naive approach to learn low-dimensional latent representations of users and items. Its simplicity lies in the fact that it solely relies on collaborative filtering signals without considering item contents or user profiles. As a result, BPRMF generally yields sub-optimal performance in user preference modeling, highlighting the insufficiency of solely modeling direct user-item interactions. For LightGCN, as we expected, it consistently outperforms BPRMF, verifying that stacking GCN layers is an effective way to model high-order relationships which is vital to capture user preference. Moreover, in some of the metrics, LightGCN can reach better performance than TPR, this again highlights the importance of high-order connectivity modeling. Nevertheless, it still relies on the information without any user profile and item contents. RGCL, the state-of-the-art review-based recommendation model in rating prediction, utilizes the user reviews and GCN structure to model high-order connectivity for users and items; however, it performs terribly worse in top-N tasks, probably not only because of the intrinsic gap between rating predictions and top-N tasks, but also the exclusion of content information such as titles and images.

TPR generally has competitive results with the LightGCN. In the Grocery dataset with Recall@5 and Recall@10, TPR performs significantly better than LightGCN. We suggest this could be that the item textual contents in Grocery data tend to contain much more information, which assists the model in capturing additional collaborative filtering signals. This again proves the importance of modeling item textual contents for representation learning. However, their tremendous KG structure, i.e., treating word tokens, items, and users as different nodes in the KG, makes them unable to learn information effectively when the token-level information is not enough. 

AGTM surpasses the preceding models due to its utilization of item modeling that incorporates textual contents at the sentence-level, resulting in improved performance compared to TPR. AGTM effectively captures high-order connectivity through the use of a GCN-based model, further highlighting the potency of GCN in recommendation tasks. However, in their model design, AGTM does not directly train user embeddings. Instead, the user is modeled using an aspect attentive network. As a consequence, the user embedding does not undergo direct training during the backward propagation of the model, resulting in information loss and a deficiency in learning representations. 

Lastly, \printf model consistently achieves exceptional performance across all datasets, providing strong evidence for the effectiveness of our model design. The observed improvements are statistically significant, with all $p$-values being less than 0.01. These improvements can be attributed to three key factors: 
\vspace{-12pt}
\begin{itemize}[leftmargin=*]
    \item The cross-modality feature alignment encoder in CMIM enables us to extract more potent and robust image and text features with global semantics. 
    \item Using dimension-based attention via extracted review representations for user embeddings enables us to effectively determine the prominence relationship between user preferences and item contents. Moreover, the user embeddings are directly trained during the EPIM stage to avoid information loss. 
    \item The bipartite graph structure in EPIM allows \printf model to successfully distill higher-order relationships and precisely model user preferences.
\end{itemize}
% (1) The cross-modality feature alignment encoder in CMIM enables us to extract more potent and robust image and text features with global semantics. (2) Using dimension-based attention via extracted review representations for user embeddings enables us to effectively determine the prominence relationship between user preferences and item contents. Moreover, the user embeddings are directly trained during the EPIM stage to avoid information loss. (3) The bipartite graph structure in EPIM allows \printf model to successfully distill higher-order relationships and precisely model user preferences.

\subsection{Ablation Study}

\label{sec:ablation_studies}
\begin{table*}[t!]
\resizebox{0.99\textwidth}{!}{
\begin{tabular}[t]{l|cc|cc|cc|cc|cc|cc}
\toprule
 & \multicolumn{4}{c|}{\textbf{Amazon-Grocery}} & \multicolumn{4}{c|}{\textbf{Amazon-Tools}} & \multicolumn{4}{c}{\textbf{Amazon-Electronics}} \\
 & \textbf{R@5} & \textbf{N@5} & \textbf{R@10} & \textbf{N@10} & \textbf{R@5} & \textbf{N@5} & \textbf{R@10} & \textbf{N@10} & \textbf{R@5} & \textbf{N@5} & \textbf{R@10} & \textbf{N@10} \\ 
\midrule
TPR (text only) & 0.0923 & 0.0918 & 0.1228 & 0.1031 & 0.0269 & 0.0248 & 0.0405 & 0.0302 & 0.0222 & 0.0251 & 0.0362 & 0.0301 \\
TPR w/ image & 0.1077 & 0.1112 & 0.1390 & 0.1222 & 0.0346 & 0.0333 & 0.0497 & 0.0393 & 0.0284 & 0.0323 & 0.0438 & 0.0376 \\
\midrule
AGTM (text only) & 0.1001 & 0.1050 & 0.1337 & 0.1167 & 0.0336 & 0.0315 & 0.0519 & 0.0386 & 0.0309 & 0.0370 & 0.0498 & 0.0433 \\
AGTM w/ image & 0.1141 & 0.1220 & 0.1429 & 0.1312 & 0.0395 & 0.0385 & 0.0558 & 0.0448 & 0.0345 & 0.0407 & 0.0528 & 0.0466 \\
\midrule
printf w/o image & 0.0970 & 0.1025 & 0.1298 & 0.1137 & 0.0291 & 0.0273 & 0.0460 & 0.0339 & 0.0263 & 0.0310 & 0.0434 & 0.0370 \\
printf & 0.1233 & 0.1331 & 0.1536 & 0.1426 & 0.0466 & 0.0468 & 0.0631 & 0.0530 & 0.0390 & 0.0465 & 0.0574 & 0.0522 \\
\bottomrule
\end{tabular}
}
\captionsetup{justification=centering}
\caption{Performance of different methods with and without using image contents.}
\label{tab:exp-images}
\vspace{-12pt}
\end{table*}

% [這邊又再重講一次每個component在做什麼，我覺得好像有點冗長] There are two critical components in our \printf model: (1) Cross Modality Item Modeling (CMIM), which encodes the initial item embeddings via image and text contents, and (2) Review-Aware User Modeling (RAUM), which includes review features to extract the importance of different interacted items for users with the dimension-based attention mechanism. 

In our \printf model, we have incorporated two crucial components: CMIM and RAUM. We aim to demonstrate the effectiveness of each component individually; a comprehensive ablation study experiment is conducted. The results presented in Table~\ref{tab:exp-ablation} showcase the performance of our model when each component is excluded, allowing us to evaluate the impact and efficiency of CMIM and RAUM in isolation. The constructed variants are as follows:
\begin{itemize}[leftmargin=*]
    \item \textbf{w/o CMIM (image)}: We only use the item textual features, i.e. item titles, to encode the initial item representations without using image features, removing the part of information constructed from cross-modality item modeling (CMIM).
    \item \textbf{w/o CMIM (title)}: Only item image features are included to encode the initial item representations.
    \item \textbf{w/o RAUM}: The dimension-based attention in RAUM for selecting different importance of interacted items for users is removed. The user embeddings here are randomly initialized.
    \item \textbf{w/o Both}: Both CAUM and RAUM are removed, and we directly randomly initialized the user and item embeddings in the graph for training, which is equal to LightGCN.
\end{itemize}
Based on Table~\ref{tab:exp-ablation}, it's evident that every component of our model design contributes positively to the recommendation. Removing any of these components would result in a performance decrease. Our experimental findings provide validation for the effectiveness of our design, proving that incorporating both image and textual contents as initial item embeddings is a great starting point. This integration allows CMIM to proficiently capture and align the meaning of item images and titles, enhancing item representation. It's worth noting that eliminating either textual content or image content leads to a significant decline in performance, underscoring the importance of modeling both modalities. Additionally, RAUM excels in capturing user preferences. Removing this component results in a substantial drop in performance, indicating the dimension-based attention in RAUM to differentiate the significance of interacted items is an excellent approach for user preference modeling. To conclude, all of the components play a crucial role in \printf. Note that in our ablation test, we did not include the variant of removing the entire CMIM, which entails not using item textual and image content simultaneously, while still utilizing the review-aware user modeling. The reason for this omission is that when employing dimension-based attention between review embeddings and randomly initialized item embeddings, it lacks any meaningful semantics. Consequently, when conducting ablation on CMIM, we can only eliminate either the textual contents or the visual ones, but not both simultaneously.

\subsection{Analysis and Discussions}

\subsubsection{\textbf{The Effect of GCN Layers}}
\begin{figure}[t!]
    \centering
    \includegraphics[width=0.50\textwidth]{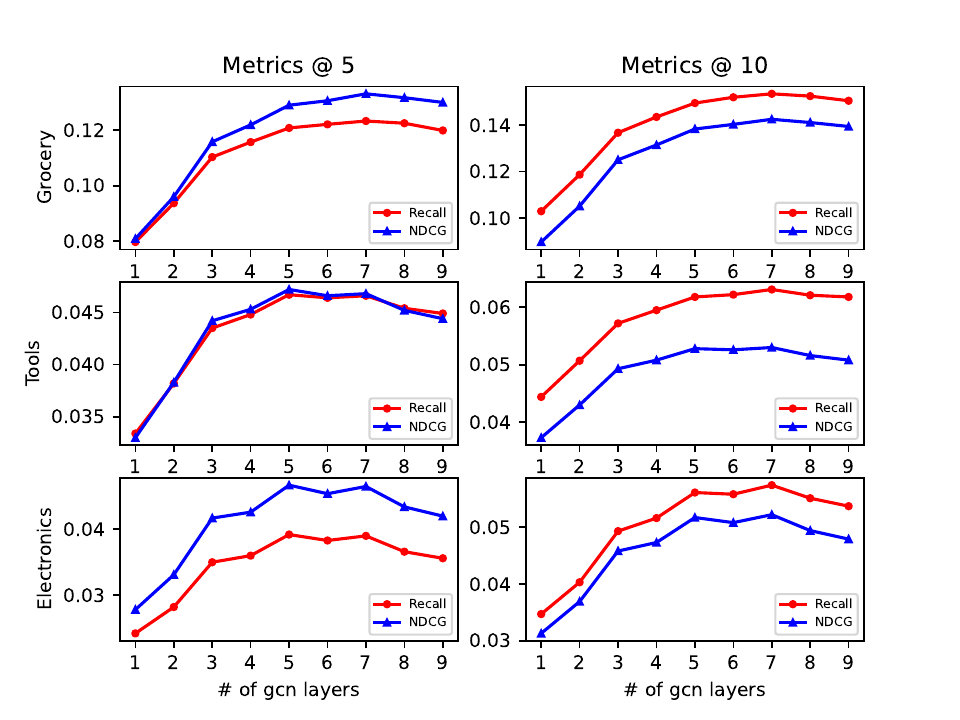}
    \caption{The effect of different numbers of GCN layers.}
    \label{fig:exp-layers}
    % \vspace{-12pt}
\end{figure}

We investigate the impact of varying the number of graph convolution layers on the performance of our model. The comprehensive experimental results can be found in Figure~\ref{fig:exp-layers}. The model performance generally improves as the number of layers is increased from 1 to 7, except for the metrics@5 on Tools and Electronics, which perform better with 5 gcn layers. In most cases, the \printf model with 7 GCN layers demonstrates competitive and satisfactory results. This finding highlights that our proposed model requires a higher number of GCN layers compared to other GCN-based models, such as LightGCN~\cite{LightGCN} and AGTM~\cite{AGTM}, which typically employ 3 GCN layers. The need for more layers in \printf may be attributed to the sparsity and diversity of textual and image features, necessitating additional layers to effectively extract user preferences and capture high-order connectivity. However, it's important to note that excessive propagation, i.e., an excessive number of GCN layers such as 8 or 9, can result in the over-smoothing of information and have a negative impact on predictions.

\begin{figure}[!t]
    \centering
    \includegraphics[width=.9\linewidth]{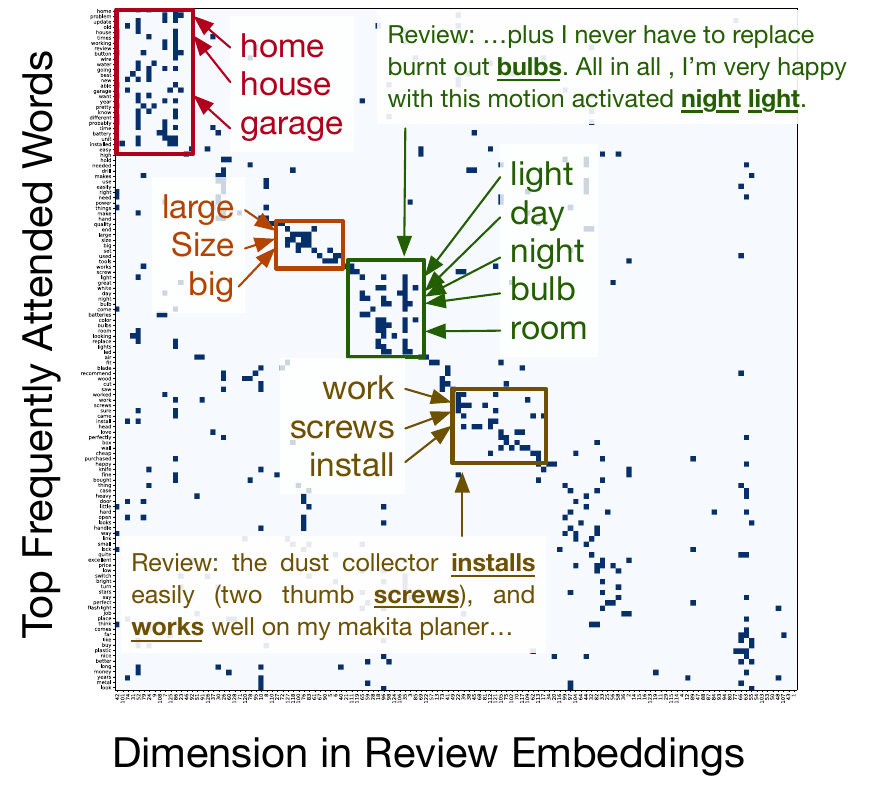}
    \caption{Bi-clusters of top frequently attended words between different review embedding dimensions derived from the Amazon-Tools dataset.}
    \label{fig:word-dim}
    \vspace{-18pt}
\end{figure}

\begin{figure*}[!]
    \centering
    \includegraphics[width=\textwidth]{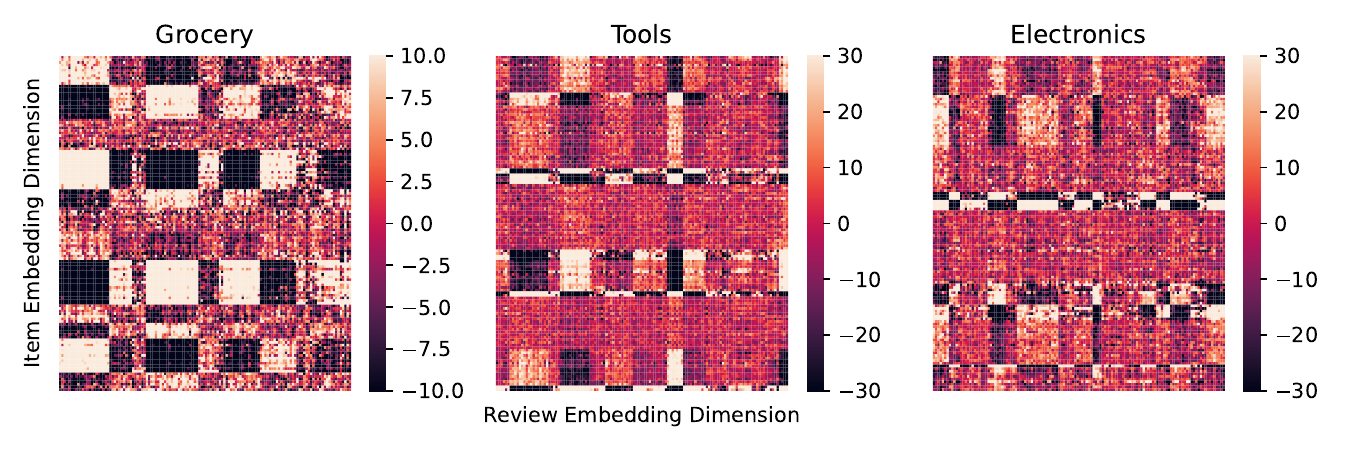}
    \caption{The direct relationship between user review embeddings and item content embeddings.}
    \label{fig:attn-matrix}
    \vspace{-6pt}
\end{figure*}

\subsubsection{\textbf{The Effect of Image Contents}}\label{sec:effect_of_images}
Although we have compared the performance with our proposed \printf model without incorporating image content in Section~\ref{sec:ablation_studies}, we would like to validate whether adding image content is a great choice for previous state-of-the-art models. Due to the constraint of model design such as TPR, we are unable to add image information toward its training phase; therefore, for those baseline models that only utilized the item textual information, we manually add the image information to their user and item representation. The most naive method is that for items, use the same method as \printf to concatenate the item image representation behind the item embedding of the original baseline model; on the other hand, for users, use the average image embedding of items he has interacted with as the user image representation, and concatenate behind the user embedding of the original baseline model, so that the prediction score can also be obtained through the dot-product as original model settings. However, this leads to negative effects and instead creates noise that interferes with the originally generated representation, since the image embedding we obtained did not participate in the training of the recommendation task. Therefore, we design a method that uses the item image representation and user image representation we just get, as the initial embedding of BPRMF, and conducts model training with the current recommendation task, then takes the user and item embeddings which have been fully trained. We concatenate the embeddings individually behind the user and item representation to predict the top-N items, and compare the performance. From the Table~\ref{tab:exp-images}, all the above models have positive effects when image contents are taken into consideration.

\subsubsection{\textbf{Dimension-Based Attention in RAUM}}
Here we would like to validate and evaluate the assumptions and designs we made in RAUM. The assumption posits that each dimension in the representation corresponds to specific topics and aspects that can be easily differentiated. To investigate this hypothesis, we leveraged the attention scores calculated for each dimension to observe if such a phenomenon exists as shown in Figure~\ref{fig:word-dim}, in which we identified the top frequently attended words for each dimension. It conclusively demonstrated that indeed, each dimension possessed words that were distinctly associated with it, as evident through their diagonal co-cluster effect. Furthermore, these words encompassed diverse types and themes. For instance, within the word cluster positioned in the middle (bound with a green rectangle and arrows pointed), the words are "light, day, night, bulb, room," it can be easily inferred that these particular dimensions pertain to items related to indoor electric lights, encompassing discussions about the brightness and color characteristics of the products.

We also explored the direct relationships between different dimensions of the review representations and item representations since we hypothesized that the text within a review would have a certain connection with the item's title or picture. To apply the concept to the embedding representation within the latent space, each dimension of the review embedding should have a specific relationship with the corresponding dimension in the item embedding. By employing the formula developed in Section~\ref{sec:dim-attn-raum}, we obtained matrix $D$ successfully. Subsequently, through spectral clustering analysis, we were able to confirm the presence of a specific correlation between the review and the embedding dimensions of the item. Referring to Figure~\ref{fig:attn-matrix}, we can initially observe the successful differentiation between review embedding dimensions and item embedding dimensions. This differentiation enables us to trace back the importance of the current review embedding dimension and its corresponding dimensions in item embeddings. By doing so, we gain insights into the frequent transmission of specific information between these two latent spaces and understand the type of information that is transparent in-between.

To conclude, for RAUM, we can treat the matrix $D$ not only as a projection for the review from its space to the item space, but also as an attention matrix to further select the significance of users' reviews toward different interacted items. The design of RAUM not only effectively assists the user preference modeling, but also increases the interpretability of attention.
\vspace{-6pt}

\section{Conclusion}
In this work, we proposed a new method to effectively model the common scenario in the recommendation. In the scenario, users click the item based on the image and text information provided and write down review comments. The \textbf{printf} jointly models the item's image and textual features and determines user preference based on review comments via a bipartite user-item interaction graph. The crucial keys in our model include not only the utilization of a cross-modality feature alignment encoder in CMIM to extract global semantics from item information but also the effectiveness of dimension-based attention in RAUM to make up a deficiency of the semantic gap, making the information have a better starting position of propagation in EPIM stage. Extensive experiments on three real-world datasets demonstrate the effectiveness, power, and significant improvement of our \printf.

\section*{Acknowledgements}
% We would like to thank the anonymous reviewers for their helpful comments. %老師叫我把這第一句槓掉ＸＤ
This work was partially sponsored by the Ministry of Science and Technology, Taiwan, under Grants 112-2221-E-002-197 and 111-2221-E-002-169.

\bibliographystyle{ACM-Reference-Format}
\balance
\bibliography{sample-base}

%%% -*-BibTeX-*-
%%% Do NOT edit. File created by BibTeX with style
%%% ACM-Reference-Format-Journals [18-Jan-2012].

\begin{thebibliography}{51}

%%% ====================================================================
%%% NOTE TO THE USER: you can override these defaults by providing
%%% customized versions of any of these macros before the \bibliography
%%% command.  Each of them MUST provide its own final punctuation,
%%% except for \shownote{}, \showDOI{}, and \showURL{}.  The latter two
%%% do not use final punctuation, in order to avoid confusing it with
%%% the Web address.
%%%
%%% To suppress output of a particular field, define its macro to expand
%%% to an empty string, or better, \unskip, like this:
%%%
%%% \newcommand{\showDOI}[1]{\unskip}   % LaTeX syntax
%%%
%%% \def \showDOI #1{\unskip}           % plain TeX syntax
%%%
%%% ====================================================================

\ifx \showCODEN    \undefined \def \showCODEN     #1{\unskip}     \fi
\ifx \showDOI      \undefined \def \showDOI       #1{#1}\fi
\ifx \showISBNx    \undefined \def \showISBNx     #1{\unskip}     \fi
\ifx \showISBNxiii \undefined \def \showISBNxiii  #1{\unskip}     \fi
\ifx \showISSN     \undefined \def \showISSN      #1{\unskip}     \fi
\ifx \showLCCN     \undefined \def \showLCCN      #1{\unskip}     \fi
\ifx \shownote     \undefined \def \shownote      #1{#1}          \fi
\ifx \showarticletitle \undefined \def \showarticletitle #1{#1}   \fi
\ifx \showURL      \undefined \def \showURL       {\relax}        \fi
% The following commands are used for tagged output and should be
% invisible to TeX
\providecommand\bibfield[2]{#2}
\providecommand\bibinfo[2]{#2}
\providecommand\natexlab[1]{#1}
\providecommand\showeprint[2][]{arXiv:#2}

\bibitem[Cao et~al\mbox{.}(2019)]%
        {KTUP}
\bibfield{author}{\bibinfo{person}{Yixin Cao}, \bibinfo{person}{Xiang Wang},
  \bibinfo{person}{Xiangnan He}, \bibinfo{person}{Zikun Hu}, {and}
  \bibinfo{person}{Tat-Seng Chua}.} \bibinfo{year}{2019}\natexlab{}.
\newblock \showarticletitle{Unifying Knowledge Graph Learning and
  Recommendation: Towards a Better Understanding of User Preferences}. In
  \bibinfo{booktitle}{\emph{WWW}}. \bibinfo{pages}{151–161}.
\newblock


\bibitem[Chen et~al\mbox{.}(2018)]%
        {NARRE}
\bibfield{author}{\bibinfo{person}{Chong Chen}, \bibinfo{person}{Min Zhang},
  \bibinfo{person}{Yiqun Liu}, {and} \bibinfo{person}{Shaoping Ma}.}
  \bibinfo{year}{2018}\natexlab{}.
\newblock \showarticletitle{Neural Attentional Rating Regression with
  Review-Level Explanations}. In \bibinfo{booktitle}{\emph{WWW}}.
  \bibinfo{pages}{1583–1592}.
\newblock


\bibitem[Chen(2015)]%
        {TextCNN}
\bibfield{author}{\bibinfo{person}{Yahui Chen}.}
  \bibinfo{year}{2015}\natexlab{}.
\newblock \emph{\bibinfo{title}{Convolutional neural network for sentence
  classification}}.
\newblock \bibinfo{thesistype}{Master's\ thesis}. \bibinfo{school}{University
  of Waterloo}.
\newblock


\bibitem[Chen et~al\mbox{.}(2020)]%
        {UNITER}
\bibfield{author}{\bibinfo{person}{Yen-Chun Chen}, \bibinfo{person}{Linjie Li},
  \bibinfo{person}{Licheng Yu}, \bibinfo{person}{Ahmed~El Kholy},
  \bibinfo{person}{Faisal Ahmed}, \bibinfo{person}{Zhe Gan},
  \bibinfo{person}{Yu Cheng}, {and} \bibinfo{person}{Jingjing Liu}.}
  \bibinfo{year}{2020}\natexlab{}.
\newblock \bibinfo{title}{UNITER: UNiversal Image-TExt Representation
  Learning}.
\newblock
\newblock
\showeprint[arxiv]{1909.11740}~[cs.CV]


\bibitem[Chin et~al\mbox{.}(2018)]%
        {ANR}
\bibfield{author}{\bibinfo{person}{Jin~Yao Chin}, \bibinfo{person}{Kaiqi Zhao},
  \bibinfo{person}{Shafiq Joty}, {and} \bibinfo{person}{Gao Cong}.}
  \bibinfo{year}{2018}\natexlab{}.
\newblock \showarticletitle{ANR: Aspect-based neural recommender}. In
  \bibinfo{booktitle}{\emph{Proceedings of the 27th ACM International
  conference on information and knowledge management}}.
  \bibinfo{pages}{147--156}.
\newblock


\bibitem[Chuang et~al\mbox{.}(2020)]%
        {TPR}
\bibfield{author}{\bibinfo{person}{Yu-Neng Chuang}, \bibinfo{person}{Chih-Ming
  Chen}, \bibinfo{person}{Chuan-Ju Wang}, \bibinfo{person}{Ming-Feng Tsai},
  \bibinfo{person}{Yuan Fang}, {and} \bibinfo{person}{Ee-Peng Lim}.}
  \bibinfo{year}{2020}\natexlab{}.
\newblock \showarticletitle{TPR: Text-Aware Preference Ranking for Recommender
  Systems}. In \bibinfo{booktitle}{\emph{CIKM}}. \bibinfo{pages}{215–224}.
\newblock


\bibitem[Devlin et~al\mbox{.}(2019)]%
        {BERT}
\bibfield{author}{\bibinfo{person}{Jacob Devlin}, \bibinfo{person}{Ming-Wei
  Chang}, \bibinfo{person}{Kenton Lee}, {and} \bibinfo{person}{Kristina
  Toutanova}.} \bibinfo{year}{2019}\natexlab{}.
\newblock \showarticletitle{{BERT}: Pre-training of Deep Bidirectional
  Transformers for Language Understanding}. In
  \bibinfo{booktitle}{\emph{NAACL}}. \bibinfo{pages}{4171--4186}.
\newblock


\bibitem[Dosovitskiy et~al\mbox{.}(2021)]%
        {ViT}
\bibfield{author}{\bibinfo{person}{Alexey Dosovitskiy}, \bibinfo{person}{Lucas
  Beyer}, \bibinfo{person}{Alexander Kolesnikov}, \bibinfo{person}{Dirk
  Weissenborn}, \bibinfo{person}{Xiaohua Zhai}, \bibinfo{person}{Thomas
  Unterthiner}, \bibinfo{person}{Mostafa Dehghani}, \bibinfo{person}{Matthias
  Minderer}, \bibinfo{person}{Georg Heigold}, \bibinfo{person}{Sylvain Gelly},
  \bibinfo{person}{Jakob Uszkoreit}, {and} \bibinfo{person}{Neil Houlsby}.}
  \bibinfo{year}{2021}\natexlab{}.
\newblock \bibinfo{title}{An Image is Worth 16x16 Words: Transformers for Image
  Recognition at Scale}.
\newblock
\newblock
\showeprint[arxiv]{2010.11929}~[cs.CV]


\bibitem[Fan et~al\mbox{.}(2019)]%
        {GraphRec}
\bibfield{author}{\bibinfo{person}{Wenqi Fan}, \bibinfo{person}{Yao Ma},
  \bibinfo{person}{Qing Li}, \bibinfo{person}{Yuan He}, \bibinfo{person}{Eric
  Zhao}, \bibinfo{person}{Jiliang Tang}, {and} \bibinfo{person}{Dawei Yin}.}
  \bibinfo{year}{2019}\natexlab{}.
\newblock \showarticletitle{Graph neural networks for social recommendation}.
  In \bibinfo{booktitle}{\emph{The world wide web conference}}.
  \bibinfo{pages}{417--426}.
\newblock


\bibitem[Feng et~al\mbox{.}(2019)]%
        {AGCN}
\bibfield{author}{\bibinfo{person}{Chenyuan Feng}, \bibinfo{person}{Zuozhu
  Liu}, \bibinfo{person}{Shaowei Lin}, {and} \bibinfo{person}{Tony~Q.S. Quek}.}
  \bibinfo{year}{2019}\natexlab{}.
\newblock \showarticletitle{Attention-based Graph Convolutional Network for
  Recommendation System}. In \bibinfo{booktitle}{\emph{ICASSP 2019 - 2019 IEEE
  International Conference on Acoustics, Speech and Signal Processing
  (ICASSP)}}. \bibinfo{pages}{7560--7564}.
\newblock
\urldef\tempurl%
\url{https://doi.org/10.1109/ICASSP.2019.8683050}
\showDOI{\tempurl}


\bibitem[Guan et~al\mbox{.}(2019)]%
        {AARM}
\bibfield{author}{\bibinfo{person}{Xinyu Guan}, \bibinfo{person}{Zhiyong
  Cheng}, \bibinfo{person}{Xiangnan He}, \bibinfo{person}{Yongfeng Zhang},
  \bibinfo{person}{Zhibo Zhu}, \bibinfo{person}{Qinke Peng}, {and}
  \bibinfo{person}{Tat-Seng Chua}.} \bibinfo{year}{2019}\natexlab{}.
\newblock \showarticletitle{Attentive Aspect Modeling for Review-Aware
  Recommendation}.
\newblock \bibinfo{journal}{\emph{TOIS}}, Article \bibinfo{articleno}{28}
  (\bibinfo{year}{2019}), \bibinfo{numpages}{27}~pages.
\newblock


\bibitem[He et~al\mbox{.}(2020)]%
        {LightGCN}
\bibfield{author}{\bibinfo{person}{Xiangnan He}, \bibinfo{person}{Kuan Deng},
  \bibinfo{person}{Xiang Wang}, \bibinfo{person}{Yan Li},
  \bibinfo{person}{Yongdong Zhang}, {and} \bibinfo{person}{Meng Wang}.}
  \bibinfo{year}{2020}\natexlab{}.
\newblock \showarticletitle{LightGCN: Simplifying and Powering Graph
  Convolution Network for Recommendation}. In
  \bibinfo{booktitle}{\emph{SIGIR}}. \bibinfo{pages}{639–648}.
\newblock


\bibitem[He et~al\mbox{.}(2017)]%
        {NeuralCF}
\bibfield{author}{\bibinfo{person}{Xiangnan He}, \bibinfo{person}{Lizi Liao},
  \bibinfo{person}{Hanwang Zhang}, \bibinfo{person}{Liqiang Nie},
  \bibinfo{person}{Xia Hu}, {and} \bibinfo{person}{Tat-Seng Chua}.}
  \bibinfo{year}{2017}\natexlab{}.
\newblock \showarticletitle{Neural Collaborative Filtering}. In
  \bibinfo{booktitle}{\emph{WWW}}. \bibinfo{pages}{173–182}.
\newblock


\bibitem[Juan et~al\mbox{.}(2023)]%
        {AGTM}
\bibfield{author}{\bibinfo{person}{Ming-Hao Juan}, \bibinfo{person}{Pu-Jen
  Cheng}, \bibinfo{person}{Hui-Neng Hsu}, {and} \bibinfo{person}{Pin-Hsin
  Hsiao}.} \bibinfo{year}{2023}\natexlab{}.
\newblock \bibinfo{title}{Attentive Graph-based Text-aware Preference Modeling
  for Top-N Recommendation}.
\newblock
\newblock
\showeprint[arxiv]{2305.12976}~[cs.IR]


\bibitem[Khosla et~al\mbox{.}(2021)]%
        {Contrastive}
\bibfield{author}{\bibinfo{person}{Prannay Khosla}, \bibinfo{person}{Piotr
  Teterwak}, \bibinfo{person}{Chen Wang}, \bibinfo{person}{Aaron Sarna},
  \bibinfo{person}{Yonglong Tian}, \bibinfo{person}{Phillip Isola},
  \bibinfo{person}{Aaron Maschinot}, \bibinfo{person}{Ce Liu}, {and}
  \bibinfo{person}{Dilip Krishnan}.} \bibinfo{year}{2021}\natexlab{}.
\newblock \bibinfo{title}{Supervised Contrastive Learning}.
\newblock
\newblock
\showeprint[arxiv]{2004.11362}~[cs.LG]


\bibitem[Krizhevsky et~al\mbox{.}(2017)]%
        {ImageNetCNN}
\bibfield{author}{\bibinfo{person}{Alex Krizhevsky}, \bibinfo{person}{Ilya
  Sutskever}, {and} \bibinfo{person}{Geoffrey~E Hinton}.}
  \bibinfo{year}{2017}\natexlab{}.
\newblock \showarticletitle{Imagenet classification with deep convolutional
  neural networks}.
\newblock \bibinfo{journal}{\emph{Commun. ACM}} \bibinfo{volume}{60},
  \bibinfo{number}{6} (\bibinfo{year}{2017}), \bibinfo{pages}{84--90}.
\newblock


\bibitem[Li et~al\mbox{.}(2022)]%
        {BLIP}
\bibfield{author}{\bibinfo{person}{Junnan Li}, \bibinfo{person}{Dongxu Li},
  \bibinfo{person}{Caiming Xiong}, {and} \bibinfo{person}{Steven Hoi}.}
  \bibinfo{year}{2022}\natexlab{}.
\newblock \showarticletitle{{BLIP}: Bootstrapping Language-Image Pre-training
  for Unified Vision-Language Understanding and Generation}. In
  \bibinfo{booktitle}{\emph{Proceedings of the 39th International Conference on
  Machine Learning}} \emph{(\bibinfo{series}{Proceedings of Machine Learning
  Research}, Vol.~\bibinfo{volume}{162})},
  \bibfield{editor}{\bibinfo{person}{Kamalika Chaudhuri},
  \bibinfo{person}{Stefanie Jegelka}, \bibinfo{person}{Le~Song},
  \bibinfo{person}{Csaba Szepesvari}, \bibinfo{person}{Gang Niu}, {and}
  \bibinfo{person}{Sivan Sabato}} (Eds.). \bibinfo{publisher}{PMLR},
  \bibinfo{pages}{12888--12900}.
\newblock
\urldef\tempurl%
\url{https://proceedings.mlr.press/v162/li22n.html}
\showURL{%
\tempurl}


\bibitem[Li et~al\mbox{.}(2021)]%
        {ALBEF}
\bibfield{author}{\bibinfo{person}{Junnan Li}, \bibinfo{person}{Ramprasaath
  Selvaraju}, \bibinfo{person}{Akhilesh Gotmare}, \bibinfo{person}{Shafiq
  Joty}, \bibinfo{person}{Caiming Xiong}, {and} \bibinfo{person}{Steven
  Chu~Hong Hoi}.} \bibinfo{year}{2021}\natexlab{}.
\newblock \showarticletitle{Align before fuse: Vision and language
  representation learning with momentum distillation}.
\newblock \bibinfo{journal}{\emph{Advances in neural information processing
  systems}}  \bibinfo{volume}{34} (\bibinfo{year}{2021}),
  \bibinfo{pages}{9694--9705}.
\newblock


\bibitem[Li et~al\mbox{.}(2019)]%
        {VisualBERT}
\bibfield{author}{\bibinfo{person}{Liunian~Harold Li}, \bibinfo{person}{Mark
  Yatskar}, \bibinfo{person}{Da Yin}, \bibinfo{person}{Cho-Jui Hsieh}, {and}
  \bibinfo{person}{Kai-Wei Chang}.} \bibinfo{year}{2019}\natexlab{}.
\newblock \bibinfo{title}{VisualBERT: A Simple and Performant Baseline for
  Vision and Language}.
\newblock
\newblock
\showeprint[arxiv]{1908.03557}~[cs.CV]


\bibitem[Li and Xu(2020)]%
        {AFRAM}
\bibfield{author}{\bibinfo{person}{Weiqian Li} {and} \bibinfo{person}{Bugao
  Xu}.} \bibinfo{year}{2020}\natexlab{}.
\newblock \showarticletitle{Aspect-based fashion recommendation with attention
  mechanism}.
\newblock \bibinfo{journal}{\emph{IEEE Access}}  \bibinfo{volume}{8}
  (\bibinfo{year}{2020}), \bibinfo{pages}{141814--141823}.
\newblock


\bibitem[Linden et~al\mbox{.}(2003)]%
        {AmazonCF}
\bibfield{author}{\bibinfo{person}{Greg Linden}, \bibinfo{person}{Brent Smith},
  {and} \bibinfo{person}{Jeremy York}.} \bibinfo{year}{2003}\natexlab{}.
\newblock \showarticletitle{Amazon. com recommendations: Item-to-item
  collaborative filtering}.
\newblock \bibinfo{journal}{\emph{IEEE Internet computing}}
  \bibinfo{volume}{7}, \bibinfo{number}{1} (\bibinfo{year}{2003}),
  \bibinfo{pages}{76--80}.
\newblock


\bibitem[Liu et~al\mbox{.}(2020)]%
        {NRCA}
\bibfield{author}{\bibinfo{person}{Hongtao Liu}, \bibinfo{person}{Wenjun Wang},
  \bibinfo{person}{Hongyan Xu}, \bibinfo{person}{Qiyao Peng}, {and}
  \bibinfo{person}{Pengfei Jiao}.} \bibinfo{year}{2020}\natexlab{}.
\newblock \showarticletitle{Neural Unified Review Recommendation with Cross
  Attention}. In \bibinfo{booktitle}{\emph{SIGIR}}.
  \bibinfo{pages}{1789–1792}.
\newblock


\bibitem[Loshchilov and Hutter(2019)]%
        {AdamW}
\bibfield{author}{\bibinfo{person}{Ilya Loshchilov} {and}
  \bibinfo{person}{Frank Hutter}.} \bibinfo{year}{2019}\natexlab{}.
\newblock \showarticletitle{Decoupled Weight Decay Regularization}. In
  \bibinfo{booktitle}{\emph{ICLR}}.
\newblock


\bibitem[Lu et~al\mbox{.}(2019)]%
        {VILBERT}
\bibfield{author}{\bibinfo{person}{Jiasen Lu}, \bibinfo{person}{Dhruv Batra},
  \bibinfo{person}{Devi Parikh}, {and} \bibinfo{person}{Stefan Lee}.}
  \bibinfo{year}{2019}\natexlab{}.
\newblock \bibinfo{title}{ViLBERT: Pretraining Task-Agnostic Visiolinguistic
  Representations for Vision-and-Language Tasks}.
\newblock
\newblock
\showeprint[arxiv]{1908.02265}~[cs.CV]


\bibitem[Peters et~al\mbox{.}(2018)]%
        {ELMO}
\bibfield{author}{\bibinfo{person}{Matthew~E. Peters}, \bibinfo{person}{Mark
  Neumann}, \bibinfo{person}{Mohit Iyyer}, \bibinfo{person}{Matt Gardner},
  \bibinfo{person}{Christopher Clark}, \bibinfo{person}{Kenton Lee}, {and}
  \bibinfo{person}{Luke Zettlemoyer}.} \bibinfo{year}{2018}\natexlab{}.
\newblock \bibinfo{title}{Deep contextualized word representations}.
\newblock
\newblock
\showeprint[arxiv]{1802.05365}~[cs.CL]


\bibitem[Qiu et~al\mbox{.}(2019)]%
        {qiu2019rethinking}
\bibfield{author}{\bibinfo{person}{Ruihong Qiu}, \bibinfo{person}{Jingjing Li},
  \bibinfo{person}{Zi Huang}, {and} \bibinfo{person}{Hongzhi Yin}.}
  \bibinfo{year}{2019}\natexlab{}.
\newblock \showarticletitle{Rethinking the item order in session-based
  recommendation with graph neural networks}. In
  \bibinfo{booktitle}{\emph{Proceedings of the 28th ACM international
  conference on information and knowledge management}}.
  \bibinfo{pages}{579--588}.
\newblock


\bibitem[Radford et~al\mbox{.}(2021)]%
        {CLIP}
\bibfield{author}{\bibinfo{person}{Alec Radford}, \bibinfo{person}{Jong~Wook
  Kim}, \bibinfo{person}{Chris Hallacy}, \bibinfo{person}{Aditya Ramesh},
  \bibinfo{person}{Gabriel Goh}, \bibinfo{person}{Sandhini Agarwal},
  \bibinfo{person}{Girish Sastry}, \bibinfo{person}{Amanda Askell},
  \bibinfo{person}{Pamela Mishkin}, \bibinfo{person}{Jack Clark},
  \bibinfo{person}{Gretchen Krueger}, {and} \bibinfo{person}{Ilya Sutskever}.}
  \bibinfo{year}{2021}\natexlab{}.
\newblock \bibinfo{title}{Learning Transferable Visual Models From Natural
  Language Supervision}.
\newblock
\newblock
\showeprint[arxiv]{2103.00020}~[cs.CV]


\bibitem[Reimers and Gurevych(2019)]%
        {SBERT}
\bibfield{author}{\bibinfo{person}{Nils Reimers} {and} \bibinfo{person}{Iryna
  Gurevych}.} \bibinfo{year}{2019}\natexlab{}.
\newblock \showarticletitle{Sentence-BERT: Sentence Embeddings using Siamese
  BERT-Networks}. In \bibinfo{booktitle}{\emph{EMNLP}}.
\newblock


\bibitem[Rendle et~al\mbox{.}(2009)]%
        {BPR}
\bibfield{author}{\bibinfo{person}{Steffen Rendle}, \bibinfo{person}{Christoph
  Freudenthaler}, \bibinfo{person}{Zeno Gantner}, {and} \bibinfo{person}{Lars
  Schmidt-Thieme}.} \bibinfo{year}{2009}\natexlab{}.
\newblock \showarticletitle{BPR: Bayesian Personalized Ranking from Implicit
  Feedback}. In \bibinfo{booktitle}{\emph{UAI}}. \bibinfo{pages}{452–461}.
\newblock


\bibitem[Rumelhart et~al\mbox{.}(1986)]%
        {AE}
\bibfield{author}{\bibinfo{person}{D.~E. Rumelhart}, \bibinfo{person}{G.~E.
  Hinton}, {and} \bibinfo{person}{R.~J. Williams}.}
  \bibinfo{year}{1986}\natexlab{}.
\newblock \showarticletitle{Learning Internal Representations by Error
  Propagation}. In \bibinfo{booktitle}{\emph{Parallel Distributed Processing:
  Explorations in the Microstructure of Cognition, Vol. 1: Foundations}}.
  \bibinfo{pages}{318–362}.
\newblock


\bibitem[Sarwar et~al\mbox{.}(2001)]%
        {ItemCF}
\bibfield{author}{\bibinfo{person}{Badrul Sarwar}, \bibinfo{person}{George
  Karypis}, \bibinfo{person}{Joseph Konstan}, {and} \bibinfo{person}{John
  Riedl}.} \bibinfo{year}{2001}\natexlab{}.
\newblock \showarticletitle{Item-based collaborative filtering recommendation
  algorithms}. In \bibinfo{booktitle}{\emph{Proceedings of the 10th
  international conference on World Wide Web}}. \bibinfo{pages}{285--295}.
\newblock


\bibitem[Schafer et~al\mbox{.}(2007)]%
        {CF}
\bibfield{author}{\bibinfo{person}{J~Ben Schafer}, \bibinfo{person}{Dan
  Frankowski}, \bibinfo{person}{Jon Herlocker}, {and} \bibinfo{person}{Shilad
  Sen}.} \bibinfo{year}{2007}\natexlab{}.
\newblock \showarticletitle{Collaborative filtering recommender systems}.
\newblock \bibinfo{journal}{\emph{The adaptive web: methods and strategies of
  web personalization}} (\bibinfo{year}{2007}), \bibinfo{pages}{291--324}.
\newblock


\bibitem[Shi et~al\mbox{.}(2019)]%
        {shi2019deep}
\bibfield{author}{\bibinfo{person}{Chuan Shi}, \bibinfo{person}{Xiaotian Han},
  \bibinfo{person}{Li Song}, \bibinfo{person}{Xiao Wang},
  \bibinfo{person}{Senzhang Wang}, \bibinfo{person}{Junping Du}, {and}
  \bibinfo{person}{S~Yu Philip}.} \bibinfo{year}{2019}\natexlab{}.
\newblock \showarticletitle{Deep collaborative filtering with multi-aspect
  information in heterogeneous networks}.
\newblock \bibinfo{journal}{\emph{IEEE transactions on knowledge and data
  engineering}} \bibinfo{volume}{33}, \bibinfo{number}{4}
  (\bibinfo{year}{2019}), \bibinfo{pages}{1413--1425}.
\newblock


\bibitem[Shuai et~al\mbox{.}(2022)]%
        {RGCL}
\bibfield{author}{\bibinfo{person}{Jie Shuai}, \bibinfo{person}{Kun Zhang},
  \bibinfo{person}{Le Wu}, \bibinfo{person}{Peijie Sun},
  \bibinfo{person}{Richang Hong}, \bibinfo{person}{Meng Wang}, {and}
  \bibinfo{person}{Yong Li}.} \bibinfo{year}{2022}\natexlab{}.
\newblock \showarticletitle{A Review-aware Graph Contrastive Learning Framework
  for Recommendation}. In \bibinfo{booktitle}{\emph{Proceedings of the 45th
  International {ACM} {SIGIR} Conference on Research and Development in
  Information Retrieval}}. \bibinfo{publisher}{{ACM}}.
\newblock
\urldef\tempurl%
\url{https://doi.org/10.1145/3477495.3531927}
\showDOI{\tempurl}


\bibitem[Touvron et~al\mbox{.}(2021)]%
        {DEIT}
\bibfield{author}{\bibinfo{person}{Hugo Touvron}, \bibinfo{person}{Matthieu
  Cord}, \bibinfo{person}{Matthijs Douze}, \bibinfo{person}{Francisco Massa},
  \bibinfo{person}{Alexandre Sablayrolles}, {and} \bibinfo{person}{Herve
  Jegou}.} \bibinfo{year}{2021}\natexlab{}.
\newblock \showarticletitle{Training data-efficient image transformers
  distillation through attention}. In \bibinfo{booktitle}{\emph{International
  Conference on Machine Learning}}, Vol.~\bibinfo{volume}{139}.
  \bibinfo{pages}{10347--10357}.
\newblock


\bibitem[van~den Berg et~al\mbox{.}(2018)]%
        {GCMC}
\bibfield{author}{\bibinfo{person}{Rianne van~den Berg},
  \bibinfo{person}{Thomas~N. Kipf}, {and} \bibinfo{person}{Max Welling}.}
  \bibinfo{year}{2018}\natexlab{}.
\newblock \showarticletitle{Graph Convolutional Matrix Completion}. In
  \bibinfo{booktitle}{\emph{KDD Workshop on Deep Learning Day}}.
\newblock


\bibitem[Vaswani et~al\mbox{.}(2017)]%
        {Attention}
\bibfield{author}{\bibinfo{person}{Ashish Vaswani}, \bibinfo{person}{Noam
  Shazeer}, \bibinfo{person}{Niki Parmar}, \bibinfo{person}{Jakob Uszkoreit},
  \bibinfo{person}{Llion Jones}, \bibinfo{person}{Aidan~N Gomez},
  \bibinfo{person}{{\L}ukasz Kaiser}, {and} \bibinfo{person}{Illia
  Polosukhin}.} \bibinfo{year}{2017}\natexlab{}.
\newblock \showarticletitle{Attention is all you need}.
\newblock \bibinfo{journal}{\emph{Advances in neural information processing
  systems}}  \bibinfo{volume}{30} (\bibinfo{year}{2017}).
\newblock


\bibitem[Wang et~al\mbox{.}(2019c)]%
        {KGCN}
\bibfield{author}{\bibinfo{person}{Hongwei Wang}, \bibinfo{person}{Miao Zhao},
  \bibinfo{person}{Xing Xie}, \bibinfo{person}{Wenjie Li}, {and}
  \bibinfo{person}{Minyi Guo}.} \bibinfo{year}{2019}\natexlab{c}.
\newblock \showarticletitle{Knowledge graph convolutional networks for
  recommender systems}. In \bibinfo{booktitle}{\emph{The world wide web
  conference}}. \bibinfo{pages}{3307--3313}.
\newblock


\bibitem[Wang et~al\mbox{.}(2019a)]%
        {KGAT}
\bibfield{author}{\bibinfo{person}{Xiang Wang}, \bibinfo{person}{Xiangnan He},
  \bibinfo{person}{Yixin Cao}, \bibinfo{person}{Meng Liu}, {and}
  \bibinfo{person}{Tat-Seng Chua}.} \bibinfo{year}{2019}\natexlab{a}.
\newblock \showarticletitle{KGAT: Knowledge Graph Attention Network for
  Recommendation}. In \bibinfo{booktitle}{\emph{KDD}}.
  \bibinfo{pages}{950–958}.
\newblock


\bibitem[Wang et~al\mbox{.}(2019b)]%
        {NGCF}
\bibfield{author}{\bibinfo{person}{Xiang Wang}, \bibinfo{person}{Xiangnan He},
  \bibinfo{person}{Meng Wang}, \bibinfo{person}{Fuli Feng}, {and}
  \bibinfo{person}{Tat-Seng Chua}.} \bibinfo{year}{2019}\natexlab{b}.
\newblock \showarticletitle{Neural Graph Collaborative Filtering}. In
  \bibinfo{booktitle}{\emph{SIGIR}}. \bibinfo{pages}{165–174}.
\newblock


\bibitem[Wang et~al\mbox{.}(2021)]%
        {KGIN}
\bibfield{author}{\bibinfo{person}{Xiang Wang}, \bibinfo{person}{Tinglin
  Huang}, \bibinfo{person}{Dingxian Wang}, \bibinfo{person}{Yancheng Yuan},
  \bibinfo{person}{Zhenguang Liu}, \bibinfo{person}{Xiangnan He}, {and}
  \bibinfo{person}{Tat-Seng Chua}.} \bibinfo{year}{2021}\natexlab{}.
\newblock \showarticletitle{Learning Intents behind Interactions with Knowledge
  Graph for Recommendation}. In \bibinfo{booktitle}{\emph{WWW}}.
  \bibinfo{pages}{878–887}.
\newblock


\bibitem[Wang et~al\mbox{.}(2020)]%
        {CKAN}
\bibfield{author}{\bibinfo{person}{Ze Wang}, \bibinfo{person}{Guangyan Lin},
  \bibinfo{person}{Huobin Tan}, \bibinfo{person}{Qinghong Chen}, {and}
  \bibinfo{person}{Xiyang Liu}.} \bibinfo{year}{2020}\natexlab{}.
\newblock \showarticletitle{CKAN: Collaborative Knowledge-Aware Attentive
  Network for Recommender Systems}. In \bibinfo{booktitle}{\emph{SIGIR}}.
  \bibinfo{pages}{219–228}.
\newblock


\bibitem[Wu et~al\mbox{.}(2019a)]%
        {NAML}
\bibfield{author}{\bibinfo{person}{Chuhan Wu}, \bibinfo{person}{Fangzhao Wu},
  \bibinfo{person}{Mingxiao An}, \bibinfo{person}{Jianqiang Huang},
  \bibinfo{person}{Yongfeng Huang}, {and} \bibinfo{person}{Xing Xie}.}
  \bibinfo{year}{2019}\natexlab{a}.
\newblock \showarticletitle{Neural News Recommendation with Attentive
  Multi-View Learning}. In \bibinfo{booktitle}{\emph{IJCAI}}.
  \bibinfo{pages}{3863–3869}.
\newblock


\bibitem[Wu et~al\mbox{.}(2019b)]%
        {HUITA}
\bibfield{author}{\bibinfo{person}{Chuhan Wu}, \bibinfo{person}{Fangzhao Wu},
  \bibinfo{person}{Junxin Liu}, {and} \bibinfo{person}{Yongfeng Huang}.}
  \bibinfo{year}{2019}\natexlab{b}.
\newblock \showarticletitle{Hierarchical user and item representation with
  three-tier attention for recommendation}. In
  \bibinfo{booktitle}{\emph{Proceedings of the 2019 Conference of the North
  American Chapter of the Association for Computational Linguistics: Human
  Language Technologies, Volume 1 (Long and Short Papers)}}.
  \bibinfo{pages}{1818--1826}.
\newblock


\bibitem[Yang et~al\mbox{.}(2018)]%
        {HOPRec}
\bibfield{author}{\bibinfo{person}{Jheng-Hong Yang}, \bibinfo{person}{Chih-Ming
  Chen}, \bibinfo{person}{Chuan-Ju Wang}, {and} \bibinfo{person}{Ming-Feng
  Tsai}.} \bibinfo{year}{2018}\natexlab{}.
\newblock \showarticletitle{HOP-Rec: High-Order Proximity for Implicit
  Recommendation}. In \bibinfo{booktitle}{\emph{RecSys}}.
  \bibinfo{pages}{140–144}.
\newblock


\bibitem[Ying et~al\mbox{.}(2018)]%
        {PinSage}
\bibfield{author}{\bibinfo{person}{Rex Ying}, \bibinfo{person}{Ruining He},
  \bibinfo{person}{Kaifeng Chen}, \bibinfo{person}{Pong Eksombatchai},
  \bibinfo{person}{William~L. Hamilton}, {and} \bibinfo{person}{Jure
  Leskovec}.} \bibinfo{year}{2018}\natexlab{}.
\newblock \showarticletitle{Graph Convolutional Neural Networks for Web-Scale
  Recommender Systems}. In \bibinfo{booktitle}{\emph{KDD}}.
  \bibinfo{pages}{974–983}.
\newblock


\bibitem[Zhang et~al\mbox{.}(2016b)]%
        {CKE}
\bibfield{author}{\bibinfo{person}{Fuzheng Zhang},
  \bibinfo{person}{Nicholas~Jing Yuan}, \bibinfo{person}{Defu Lian},
  \bibinfo{person}{Xing Xie}, {and} \bibinfo{person}{Wei-Ying Ma}.}
  \bibinfo{year}{2016}\natexlab{b}.
\newblock \showarticletitle{Collaborative Knowledge Base Embedding for
  Recommender Systems}. In \bibinfo{booktitle}{\emph{KDD}}.
  \bibinfo{pages}{353–362}.
\newblock


\bibitem[Zhang et~al\mbox{.}(2016a)]%
        {DiscreteCF}
\bibfield{author}{\bibinfo{person}{Hanwang Zhang}, \bibinfo{person}{Fumin
  Shen}, \bibinfo{person}{Wei Liu}, \bibinfo{person}{Xiangnan He},
  \bibinfo{person}{Huanbo Luan}, {and} \bibinfo{person}{Tat-Seng Chua}.}
  \bibinfo{year}{2016}\natexlab{a}.
\newblock \showarticletitle{Discrete collaborative filtering}. In
  \bibinfo{booktitle}{\emph{Proceedings of the 39th International ACM SIGIR
  conference on Research and Development in Information Retrieval}}.
  \bibinfo{pages}{325--334}.
\newblock


\bibitem[Zheng et~al\mbox{.}(2017a)]%
        {zheng2017joint}
\bibfield{author}{\bibinfo{person}{Lei Zheng}, \bibinfo{person}{Vahid Noroozi},
  {and} \bibinfo{person}{Philip~S Yu}.} \bibinfo{year}{2017}\natexlab{a}.
\newblock \showarticletitle{Joint deep modeling of users and items using
  reviews for recommendation}. In \bibinfo{booktitle}{\emph{Proceedings of the
  tenth ACM international conference on web search and data mining}}.
  \bibinfo{pages}{425--434}.
\newblock


\bibitem[Zheng et~al\mbox{.}(2017b)]%
        {DeepCoNN}
\bibfield{author}{\bibinfo{person}{Lei Zheng}, \bibinfo{person}{Vahid Noroozi},
  {and} \bibinfo{person}{Philip~S. Yu}.} \bibinfo{year}{2017}\natexlab{b}.
\newblock \showarticletitle{Joint Deep Modeling of Users and Items Using
  Reviews for Recommendation}. In \bibinfo{booktitle}{\emph{WSDM}}.
  \bibinfo{pages}{425–434}.
\newblock


\bibitem[Zhou et~al\mbox{.}(2018)]%
        {DIN}
\bibfield{author}{\bibinfo{person}{Guorui Zhou}, \bibinfo{person}{Chengru
  Song}, \bibinfo{person}{Xiaoqiang Zhu}, \bibinfo{person}{Ying Fan},
  \bibinfo{person}{Han Zhu}, \bibinfo{person}{Xiao Ma},
  \bibinfo{person}{Yanghui Yan}, \bibinfo{person}{Junqi Jin},
  \bibinfo{person}{Han Li}, {and} \bibinfo{person}{Kun Gai}.}
  \bibinfo{year}{2018}\natexlab{}.
\newblock \bibinfo{title}{Deep Interest Network for Click-Through Rate
  Prediction}.
\newblock
\newblock
\showeprint[arxiv]{1706.06978}~[stat.ML]


\end{thebibliography}

\end{document}